\documentclass[camera,letterpaper,nomarginnotes,nonarrowgutter]{jpaper} 

\usepackage{balance}
\usepackage{amsmath,amssymb,amsfonts}
\usepackage{graphicx}
\usepackage{textcomp}
\usepackage{pifont}
\usepackage{booktabs}
\usepackage{glossaries} 
\usepackage{setspace}
\usepackage{listings}
\usepackage[linesnumbered]{algorithm2e}
\usepackage{siunitx} 
\usepackage{subcaption}
\usepackage{tikz}
\usepackage{amsmath}
\usepackage{bm}
\usepackage[sort,compress]{cite}
\usepackage{footnote} 
\usepackage{multirow} 
\usepackage[toc,page]{appendix}

\usepackage{datetime} 
\usepackage{fancyhdr}

\usepackage{tcolorbox}
\tcbuselibrary{theorems}

\newtcbtheorem[]{mytheo}{Invariant}%
{colback=green!5,colframe=green!35!black,fonttitle=\bfseries}{th}

\newcommand{\XLong}[0]{Shared Activation Counters}
\newcommand{\X}[0]{SAC}
\renewcommand{\XLong}[0]{\agy{All-Bank Activation Counters for Scalable and low overhead RowHammer mitigation}}
\renewcommand{\X}[0]{\agy{ABACuS}}

\definecolor{gfored}{rgb}{0.580, 0.050, 0.211}
\definecolor{ao}{rgb}{0.007, 0.520, 0.867}
\definecolor{moegi}{rgb}{0.357, 0.537, 0.188}
\definecolor{jl}{rgb}{1.0, 0.2, 0.8}
\definecolor{brown(web)}{rgb}{0.65, 0.16, 0.16}
\definecolor{bisque}{rgb}{1.0, 0.89, 0.77}
\definecolor{nbs}{rgb}{0.88, 0.07, 0.37}
\definecolor{yt}{rgb}{0.58, 0.44, 0.86}
\definecolor{iy}{rgb}{0.0, 0.26, 0.15}

\newcommand{\dingOne}{\ding{182}}
\newcommand{\dingTwo}{\ding{183}}
\newcommand{\dingThree}{\ding{184}}
\newcommand{\dingFour}{\ding{185}}

\newif\ifdraft
\newif\ifrevision
\newif\ifcameraiter
\draftfalse
\revisionfalse
\cameraiterfalse

\newcommand{\ignore}[1]{}

\ifdraft
    \usepackage[colorinlistoftodos,prependcaption,textsize=small]{todonotes}
    \paperwidth=\dimexpr \paperwidth + 4cm\relax
    \oddsidemargin=\dimexpr\oddsidemargin + 2cm\relax
    \evensidemargin=\dimexpr\evensidemargin + 2cm\relax
    \marginparwidth=\dimexpr \marginparwidth + 2cm\relax

    \newcommand{\agy}[1]{\textcolor{gfored}{#1}}
    \newcommand{\agycomment}[1]{\todo[size=\scriptsize, linecolor=orange, bordercolor=orange, backgroundcolor=white]{\textcolor{gfored}{\textbf{@gy:} #1}}}
    \newcommand{\agyinline}[1]{{\textcolor{gfored}{\textbf{[@gy:} #1\textbf{]}}}}

    \newcommand{\atb}[1]{\textcolor{ao}{#1}}
    
    \newcommand{\atbcomment}[1]{\todo[size=\scriptsize, linecolor=orange, bordercolor=orange, backgroundcolor=white]{\textcolor{ao}{\textbf{@atb:} #1}}}
    
    \newcommand{\yct}[1]{\textcolor{yt}{#1}}
    \newcommand{\yctcomment}[1]{\todo[size=\scriptsize, linecolor=orange, bordercolor=orange, backgroundcolor=white]{\textcolor{yt}{\textbf{@yct:} #1}}}

    \newcommand{\gf}[1]{\textcolor{blue}{#1}}
    \newcommand{\gfcomment}[1]{\todo[size=\scriptsize, linecolor=orange, bordercolor=orange, backgroundcolor=white]{\textcolor{blue}{\textbf{@gf:} #1}}}
    
    \newcommand{\nb}[1]{\textcolor{nbs}{#1}}
    \newcommand{\nbcomment}[1]{\todo[size=\scriptsize, linecolor=orange, bordercolor=orange, backgroundcolor=white]{\textcolor{nbs}{\textbf{@nb:} #1}}}
    
    \newcommand{\hluo}[1]{\textcolor{moegi}{#1}}
    \newcommand{\hluocomment}[1]{\todo[size=\scriptsize, linecolor=orange, bordercolor=orange, backgroundcolor=white]{\textcolor{moegi}{\textbf{@hluo:} #1}}}

    \newcommand{\ste}[1]{\textcolor{teal}{#1}}
    \newcommand{\stecomment}[1]{\todo[size=\scriptsize, linecolor=orange, bordercolor=orange, backgroundcolor=white]{\textcolor{teal}{\textbf{@ste:} #1}}}
    \newcommand{\steinline}[1]{{\textcolor{teal}{\textbf{[@ste:} #1\textbf{]}}}}
    
    \newcommand{\iey}[1]{\textcolor{iy}{#1}}
    \newcommand{\ieycomment}[1]{\todo[size=\scriptsize, linecolor=orange, bordercolor=orange, backgroundcolor=white]{\textcolor{iy}{\textbf{@iey:} #1}}}

    \newcommand{\param}[1]{\textcolor{red}{#1}} 
\fi
\ifrevision
    \marginparwidth=\dimexpr \marginparwidth + 1cm\relax
    \usepackage[colorinlistoftodos,prependcaption,textsize=small]{todonotes}

    \newcommand{\revcommon}[1]{\textcolor{blue}{#1}}
    \definecolor{goodgreen}{rgb}{0.0, 0.5, 0.0}
    \newcommand{\revi}[1]{\textcolor{goodgreen}{#1}}
    \newcommand{\revii}[1]{\textcolor{red}{#1}}
    \newcommand{\reviii}[1]{\textcolor{purple}{#1}}

    \newcommand{\revv}[1]{\textcolor{orange}{#1}}

    \usepackage{enumitem}
    \usepackage{xargs}                    
    \usepackage{titlesec}
    \usepackage{titletoc}
    \usepackage{clipboard}
    \usepackage[all]{nowidow}
    \usepackage[framemethod=tikz]{mdframed}
    \usepackage{marginnote}
    \tcbuselibrary{breakable}
    \tcbuselibrary{hooks}
    \definecolor{lightblue}{rgb}{0.980, 0.956, 0.623}

    \newcommandx{\changev}[2][1=]{\todo[linecolor=blue,backgroundcolor=blue!25,bordercolor=blue,#1,size=\scriptsize]{#2}}
    
    \let\oldmarginnote\marginnote
    
    \renewcommand{\marginnote}[2][rectangle,draw,fill=blue!40,rounded corners]{%
            \oldmarginnote{%
            \tikz \node at (0,0) [#1]{#2};}%
            }
            
    \definecolor{lightyellow}{rgb}{0.980, 0.956, 0.623}

    \newcommand{\boxbegin} {
    	\begin{tcolorbox}[enhanced, frame hidden, colback=gray!50, breakable]
    }
    
    \newcommand{\boxend} {
    	\end{tcolorbox}
    }
    
    \newcommand{\yboxbegin} {
    	\begin{tcolorbox}[breakable, enhanced, frame hidden, colback=yellow!50]
    }
    
    \newcommand{\yboxend} {
    	\end{tcolorbox}
    }

    \mdfdefinestyle{graybox}{
        splittopskip=0,%
        splitbottomskip=0,%
        frametitleaboveskip=0,
        frametitlebelowskip=0,
        skipabove=0,%
        skipbelow=0,%
        leftmargin=0,%
        rightmargin=0,%
        innertopmargin=0mm,%
        innerbottommargin=0mm,%
        roundcorner=1mm,%
        backgroundcolor=lightblue,
        hidealllines=true}
        
    \mdfdefinestyle{graybox2}{
        splittopskip=0,%
        splitbottomskip=0,%
        frametitleaboveskip=0,
        frametitlebelowskip=0,
        skipabove=0,%
        skipbelow=0,%
        leftmargin=0,%
        rightmargin=0,%
        innertopmargin=2mm,%
        innerbottommargin=2mm,%
        roundcorner=2mm,%
        backgroundcolor=lightblue,
        hidealllines=true}
    
    \newcommand{\bboxbegin}{
        \begin{mdframed}[style=graybox]
    }
    
    \newcommand{\bboxend}{
        \end{mdframed}
    }
    
    \newcommand{\yyboxbegin}{
        \begin{mdframed}[style=graybox2]
    }
    
    \newcommand{\yyboxend}{
        \end{mdframed}
    }

    \newcommand{\agy}[1]{{#1}}
    \newcommand{\agycomment}[1]{}
    \newcommand{\agyinline}[1]{}

    \newcommand{\atb}[1]{{#1}}
    \newcommand{\atbcomment}[1]{\todo[size=\scriptsize, linecolor=orange, bordercolor=orange, backgroundcolor=white]{\textcolor{ao}{\textbf{@atb:} #1}}}
    
    \newcommand{\yct}[1]{{#1}}
    \newcommand{\yctcomment}[1]{}
    
    \newcommand{\gf}[1]{{#1}}
    \newcommand{\gfcomment}[1]{}

    \newcommand{\hluo}[1]{{#1}}
    \newcommand{\hluocomment}[1]{}
    
    \newcommand{\nb}[1]{{#1}}
    \newcommand{\nbcomment}[1]{}

    \newcommand{\iey}[1]{#1}
    \newcommand{\ieycomment}[1]{}
    
    \newcommand{\ste}[1]{#1}
    \newcommand{\stecomment}[1]{}
    \newcommand{\steinline}[1]{}

    \newcommand{\param}[1]{#1} 

\else 
    \usepackage{clipboard}
    
    \newcommand{\revcommon}[1]{{#1}}
    \definecolor{goodgreen}{rgb}{0.0, 0.5, 0.0}
    \newcommand{\revi}[1]{#1}
    \newcommand{\revii}[1]{#1}
    \newcommand{\reviii}[1]{#1}

    \newcommand{\revv}[1]{#1}
    \newcommand{\changev}[2][1=]{}

    \newcommand{\agy}[1]{{#1}}
    \newcommand{\agycomment}[1]{}
    \newcommand{\agyinline}[1]{}

    \newcommand{\atb}[1]{{#1}}
    \newcommand{\atbcomment}[1]{}
    
    \newcommand{\yct}[1]{{#1}}
    \newcommand{\yctcomment}[1]{}
    
    \newcommand{\gf}[1]{{#1}}
    \newcommand{\gfcomment}[1]{}

    \newcommand{\hluo}[1]{{#1}}
    \newcommand{\hluocomment}[1]{}
    
    \newcommand{\nb}[1]{{#1}}
    \newcommand{\nbcomment}[1]{}

    \newcommand{\iey}[1]{#1}
    \newcommand{\ieycomment}[1]{}
    
    \newcommand{\ste}[1]{#1}
    \newcommand{\stecomment}[1]{}
    \newcommand{\steinline}[1]{}

    \newcommand{\param}[1]{{#1}} 

    \newcommand{\omcri}[1]{#1}
    \newcommand{\atbcr}[1]{#1}

\fi

\ifcameraiter
    \paperwidth=\dimexpr \paperwidth + 5cm\relax
    \oddsidemargin=\dimexpr\oddsidemargin + 2cm\relax
    \evensidemargin=\dimexpr\evensidemargin + 2cm\relax
    \marginparwidth=\dimexpr \marginparwidth + 3cm\relax
    \usepackage[colorinlistoftodos,prependcaption,textsize=small]{todonotes}
    \ifdefined\revcommon
    \else
    \newcommand{\revcommon}[1]{{#1}}
    \fi

    \renewcommand{\omcri}[1]{\textcolor{blue}{#1}}
    \renewcommand{\atbcr}[1]{\textcolor{ao}{#1}}
    \renewcommand{\atbcomment}[1]{\todo[size=\scriptsize, linecolor=orange, bordercolor=orange, backgroundcolor=white]{\textcolor{ao}{\textbf{@atb:} #1}}}

\fi

\newcommand{\ext}[1]{{#1}}

\newcounter{obs}
\newcounter{tkw}
\newcommand{\exploitingRowHammerAllCitations}[0]{\cite{fournaris2017exploiting, poddebniak2018attacking, tatar2018throwhammer, carre2018openssl, barenghi2018software, zhang2018triggering, bhattacharya2018advanced, kim2014flipping, rowhammergithub, seaborn2015exploiting, van2016drammer, gruss2016rowhammer, razavi2016flip, pessl2016drama, xiao2016one, bosman2016dedup, bhattacharya2016curious, burleson2016invited, qiao2016new, brasser2017can, jang2017sgx, aga2017good, mutlu2017rowhammer, tatar2018defeating, gruss2018another, lipp2018nethammer, van2018guardion, frigo2018grand, cojocar2019eccploit,  ji2019pinpoint, mutlu2019rowhammer, hong2019terminal, kwong2020rambleed, frigo2020trrespass, cojocar2020rowhammer, weissman2020jackhammer, zhang2020pthammer, yao2020deephammer, deridder2021smash, hassan2021utrr, jattke2022blacksmith, tol2022toward, kogler2022half, orosa2022spyhammer, zhang2022implicit, liu2022generating, cohen2022hammerscope, zheng2022trojvit, fahr2022frodo, tobah2022spechammer, rakin2022deepsteal, aydin2022cyber, mus2022jolt, wang2022research, lefforge2023reverse,fahr2022effects, kaur2022work, cai2022feasibility, li2022cyberradar, roohi2022efficient, staudigl2022neurohammer, yang2022socially, islam2022signature,kaur2023flipping,adiletta2023mayhem,dong2023onebit,liu2023hyperattack,tol2023dont}}

\newcommand{\mitigatingRowHammerAllCitations}[0]{\cite{AppleRefInc, rh-hp,rh-lenovo,greenfield2012throttling, kim2014flipping, kim2015architectural, bains14d, bains14c, bains14a, bains14b, aweke2016anvil, bains2015row, bains2016row, bains2016distributed, son2017making, seyedzadeh2018cbt,irazoqui2016mascat, you2019mrloc, lee2019twice, park2020graphene, yaglikci2021security, yaglikci2021blockhammer, frigo2020trrespass, kang2020cattwo, hassan2021utrr, qureshi2022hydra, saileshwar2022randomized, brasser2017can, konoth2018zebram, van2018guardion, vig2018rapid,  kim2022mithril, lee2021cryoguard, marazzi2022protrr, zhang2022softtrr, joardar2022learning, juffinger2023csi, yaglikci2022hira, saxena2022aqua, enomoto2022efficient, manzhosov2022revisiting, ajorpaz2022evax, naseredini2022alarm, joardar2022machine, hassan2022case, zhang2020leveraging,loughlin2021stop, devaux2021method, fakhrzadehgan2022safeguard, saroiu2022price, loughlin2022moesiprime, han2021surround, mutlu2022fundamentally, woo2022scalable, bock2019riprh, kim2015architectural, wang2021discreet, bennett2021panopticon,kim2023ddr5}}

\newcommand{\mitigatingRowHammerCounterCitations}[0]{\cite{kim2014flipping, seyedzadeh2018cbt, park2020graphene, qureshi2022hydra, kim2022mithril, marazzi2022protrr, yaglikci2022hira, loughlin2021stop,bennett2021panopticon,saxena2022aqua,saileshwar2022randomized,kim2023ddr5}}

\newcommand{\rowHammerGetsWorseCitations}[0]{\cite{kim2020revisiting, frigo2020trrespass, yaglikci2022understanding, orosa2021deeper, mutlu2017rowhammer, mutlu2018rowhammer, mutlu2019rowhammer, mutlu2022fundamentally}}

\newcommand{\figref}[1]{Fig.~\ref{#1}}

\newcommand{\secref}[1]{§\ref{#1}}

\newcommand{\nrh}[0]{N_{RH}}
\newcommand{\mech}[0]{\X{}}
\newcommand{\pprevref}[0]{p_{REF}}
\newcommand{\pnorefinaburst}[0]{p_{B}}
\newcommand{\cboost}[0]{c_B}
\newcommand{\creduc}[0]{c_R}
\newcommand{\pmin}[0]{p_{min}}
\newcommand{\pmax}[0]{p_{max}}

\newcommand{\pnorefuntilxy}[0]{P_{x,y}}
\newcommand{\pij}[0]{p_{i,j}}
\newcommand{\psa}[0]{p_{sa}}
\newcommand{\pfa}[0]{p_{fa}}
\newcommand{\pattack}[0]{p_{attack}}

\newcommand{\psecure}[0]{p_{security}}
\newcommand{\timeout}[0]{t_{to}}

\newcommand{\trcd}[0]{t_{RCD}}
\newcommand{\tras}[0]{t_{RAS}}
\newcommand{\trp}[0]{t_{RP}}
\newcommand{\trc}[0]{t_{RC}}
\newcommand{\trefi}[0]{t_{REFI}}
\newcommand{\trefw}[0]{t_{REFW}}
\newcommand{\trfc}[0]{t_{RFC}}
\newcommand{\trrd}[0]{t_{RRD}}

\newcommand{\act}[0]{ACT}
\newcommand{\pre}[0]{PRE}
\newcommand{\refresh}[0]{REF}
\newcommand{\writecmd}[0]{WR}
\newcommand{\rd}[0]{RD}

\newacronym{iqr}{$IQR$}{inter-quartile range}
\newacronym{act}{$\act{}$}{activate}
\newacronym{pre}{$\pre{}$}{precharge}
\newacronym{ref}{$\refresh{}$}{refresh}
\newacronym{wr}{$\writecmd{}$}{write}
\newacronym{rd}{$\rd{}$}{read}
\newacronym{nrh}{$\nrh$}{RowHammer threshold}
\newacronym{mech}{\emph{\X{}}}{\XLong}
\newacronym{pprevref}{$\pprevref{}$}{preventive refresh probability}
\newacronym{pnorefinaburst}{$\pnorefinaburst$}{the probability of experiencing no preventive refresh during a burst}
\newacronym{pnorefuntilxy}{$\pnorefuntilxy$}{the probability of experiencing \emph{no} preventive refresh until a given $x_{th}$ row activation in $y_{th}$ burst of activations}
\newacronym{cboost}{$\cboost{}$}{the coefficient of boosting preventive refresh probability}
\newacronym{creduc}{$\creduc{}$}{the coefficient of reducing preventive refresh probability}
\newacronym{pij}{$\pij{}$}{the probability of experiencing no preventive refresh after the row activation $i$ at a given batch $j$}
\newacronym{psa}{$\psa{}$}{the probability of a successful attempt}
\newacronym{pfa}{$\pfa{}$}{the probability of a failed attempt}
\newacronym{pattack}{$\pattack{}$}{the probability of a successful attack}
\newacronym{psecure}{$\psecure{}$}{\gls{mech}'s security guarantee over a time period, T,}
\newacronym{trefw}{$\trefw$}{refresh window}
\newacronym{trcd}{$\trcd$}{\agyinline{trcd definition here}}
\newacronym{tras}{$\tras$}{the latency of fully restoring a DRAM cell's charge}
\newacronym{trp}{$\trp$}{\agyinline{trp definition here}}
\newacronym{trc}{$\trc$}{the minimum time needed between two consecutive row activations targeting the same bank}
\newacronym{trefi}{$\trefi$}{refresh interval}
\newacronym{trfc}{$\trfc$}{refresh latency}
\newacronym{trrd}{$\trrd$}{the minimum time needed between two consecutive row activations targeting the same rank}
\newacronym{timeout}{$\timeout$}{timeout period}
\newacronym{pmin}{$\pmin$}{\agyinline{DEFINE PMIN PLEASE}}
\newacronym{pmax}{$\pmax$}{\agyinline{DEFINE PMAX PLEASE}}

\def\BibTeX{{\rm B\kern-.05em{\sc i\kern-.025em b}\kern-.08em
    T\kern-.1667em\lower.7ex\hbox{E}\kern-.125emX}}

\makeatletter
\g@addto@macro{\normalsize}{%
  \setlength{\abovedisplayskip}{2pt plus 1pt minus 1pt}
  \setlength{\belowdisplayskip}{2pt plus 1pt minus 1pt}
  \setlength{\abovedisplayshortskip}{0pt}
  \setlength{\belowdisplayshortskip}{0pt}
  \setlength{\intextsep}{2pt plus 1pt minus 1pt}
  \setlength{\textfloatsep}{4pt plus 1pt minus 1pt}
  \setlength{\skip\footins}{5pt plus 1pt minus 1pt}}
\makeatother

\usepackage{titlesec}
\titlespacing\section{0pt}{2pt plus 1pt minus 1pt}{3pt plus 1pt minus 2pt}
\titlespacing\subsection{0pt}{2pt plus 1pt minus 1pt}{3pt plus 1pt minus 2pt}
\titlespacing\subsubsection{0pt}{2pt plus 1pt minus 1pt}{3pt plus 1pt minus 2pt}
\title{\LARGE{ABACuS: All-Bank Activation Counters\\ for Scalable and Low Overhead RowHammer Mitigation}}

\usepackage[available,functional,reproduced]{usenixbadges}

\begin{document}

\author{
{Ataberk Olgun}\qquad
{Yahya Can Tugrul}\qquad
{Nisa Bostanci}\qquad
{Ismail Emir Yuksel}\qquad\\
{Haocong Luo}\and
{Steve Rhyner}\and
{Abdullah Giray Yaglikci}\and
{Geraldo F. Oliveira}\and
{Onur Mutlu}
\and\and\\\\
ETH Zurich
} 

\maketitle


\def\parsepdfdatetime#1:#2#3#4#5#6#7#8#9{%
  \def\theyear{#2#3#4#5}%
  \def\themonth{#6#7}%
  \def\theday{#8#9}%
  \parsepdftime
}

\def\parsepdftime#1#2#3#4#5#6#7\endparsepdfdatetime{%
  \def\thehour{#1#2}%
  \def\theminute{#3#4}%
  \def\thesecond{#5#6}%
  \ifstrequal{#7}{Z}
  {%
    \def\thetimezonehour{+00}%
    \def\thetimezoneminute{00}%
  }%
  {%
    \parsepdftimezone#7%
  }%
}

\def\parsepdftimezone#1'#2'{%
  \def\thetimezonehour{#1}%
  \def\thetimezoneminute{#2}%
}

\newcommand*{\thetimezone}{\thetimezonehour:\thetimezoneminute}
\expandafter\parsepdfdatetime\pdfcreationdate\endparsepdfdatetime

\settimeformat{ampmtime}
\newcommand{\versionnum}[0]{1.1}
\newcommand{\version}[1]{\emph{Version #1 (Built:~\today~@ \currenttime~UTC\thetimezone)}}

\fancyhead{}
\thispagestyle{empty}
\pagestyle{empty}
\fancyhead[C]{\textcolor{blue}{\version{\versionnum}}}
\fancypagestyle{firststyle}
{
  \fancyhead[C]{\textcolor{blue}{\version{\versionnum}}}
  \fancyfoot[C]{\thepage}
}
\thispagestyle{firststyle}
\pagestyle{plain}

\thispagestyle{plain}
\pagestyle{plain}

\vspace{-5em}
\begin{abstract}

We introduce \X{}, a new low-cost hardware-counter-based RowHammer mitigation technique that performance-, energy-, and area-efficiently scales with worsening RowHammer vulnerability. We observe that \omcri{both} \agy{benign} workloads and RowHammer attacks \agy{tend to} access DRAM rows with the same row address in multiple DRAM banks at around the same time. \agy{Based on this observation,} \X{}'s key idea is to {use a \emph{single shared row activation counter}} {to track activations to} the rows with the same row address in all DRAM banks. \agy{Unlike state-of-the-art RowHammer mitigation mechanisms that implement {a separate row activation counter} for each DRAM bank, \X{} implements fewer counters (e.g., \emph{only} one) to track an equal number of aggressor rows.}

Our comprehensive evaluations show that \X{} securely prevents RowHammer bitflips at low \agy{performance/energy} overhead and low \agy{area} cost. 
{We compare \X{} to four state-of-the-art mitigation mechanisms.}
At a near-future RowHammer threshold of 1000, \X{} incurs \emph{only} \param{0.58}\% {(0.77\%)} performance and \param{1.66}\% {(2.12\%)} DRAM energy overheads{, averaged} across \reviii{\param{62}} {single-core (8-core)} workloads, requiring \emph{only} \param{9.47} KiB of storage per DRAM rank.
At the RowHammer threshold of 1000, the best prior low-area-cost mitigation mechanism incurs \param{1.80}\% higher average performance overhead than \X{}, while \X{} requires \param{2.50}$\times{}$ smaller {chip} area to implement.
At a future RowHammer threshold of \param{125}, \X{}~performs very similarly to ({within \param{0.38\%} of the performance of}) the best {prior} performance- and energy-efficient RowHammer mitigation mechanism while requiring \param{22.72}$\times{}$ smaller chip area.
We also show that \X{}'s performance scales well with the number of DRAM banks. 
At {the} RowHammer threshold {of 125}, \X{} incurs \reviii{\param{1.58}}\%, \reviii{\param{1.50}}\%, and \reviii{\param{2.60}}\% performance overheads for 16-, 32-, and 64-bank systems across all \reviii{\param{62}} single-core workloads, respectively. \omcri{\X{} is freely and openly available at \url{https://github.com/CMU-SAFARI/ABACuS}.}

\end{abstract}
\section{Introduction}
\label{sec:introduction}

Modern DRAM chips are vulnerable to RowHammer~\cite{kim2014flipping, mutlu2017rowhammer, yang2019trap, mutlu2019rowhammer,park2016statistical, park2016experiments,
walker2021ondramrowhammer, ryu2017overcoming, yang2016suppression, yang2017scanning, gautam2019row, jiang2021quantifying,mutlu2022fundamentally-DAC}, where repeatedly opening and closing (i.e., activating and precharging, or simply {\em hammering}) a DRAM row (aggressor row) at a high enough rate can cause bitflips in physically nearby rows (victim rows). DRAM chips become more vulnerable to RowHammer as DRAM storage density increases across DRAM generations~\rowHammerGetsWorseCitations{}. The minimum number of row activations needed to induce a RowHammer bitflip, i.e., the \gls{nrh}, has reduced by more than an order of magnitude in less than a decade~\cite{kim2020revisiting}.\footnote{For example,~{RowHammer threshold (}\gls{nrh}) is \emph{only} 4.8K and 10K for some newer LPDDR4 and DDR4 DRAM chips (manufactured in 2019--2020), which is $14.4\times$ and $6.9\times$ lower than the \gls{nrh} of 69.2K for some older DRAM chips (manufactured in 2010--2013)~\cite{kim2020revisiting}.} As many prior works demonstrate on real systems~\exploitingRowHammerAllCitations{}, RowHammer bitflips can lead to security exploits that 1)~take over a system, 2)~leak security-critical or private data, and 3)~manipulate safety-critical applications' behavior {in undesirable ways}. As a result, a large body of work~\mitigatingRowHammerAllCitations{} proposes mitigation mechanisms to prevent RowHammer bitflips.

\agy{\textbf{Key Problem.}}
\agy{Many prior works {(e.g.,~\mitigatingRowHammerCounterCitations{})} propose using a set of counters to track the activation counts of potential aggressor rows (counter-based mechanisms). 
\revcommon{Using counters} {to determine rows that reach \emph{close to RowHammer thresholds} and taking mitigating actions accordingly} can prevent RowHammer bitflips at low performance and energy overheads. Unfortunately, \revcommon{mitigation mechanisms that rely on counters} face two scalability challenges. First, they need to implement an increasingly large number of counters to track all potential aggressor rows as \gls{nrh} reduces.
This is because an attacker can concurrently hammer {\emph{more}} DRAM rows when \gls{nrh} is smaller. Second, the area overhead of these mechanisms linearly increases with the number of DRAM banks in the system,
\agy{{and} modern systems {continue to use more} banks to scale up both DRAM capacity and bandwidth~\cite{salp, chang2014improving, mutlu2008parbs, subramanian2016bliss, subramanian2014bliss, lee2009improving, ebrahimi2012fairness, ipek2008self, kim2010atlas, kim2010thread, ebrahimi2011parallel, liu2012raidr, lee2013tiered, mutlu2013memory, ibmpower, amdepyc9004, intelxeonw3400}.}}
\agy{A small set of prior works (e.g.,~\cite{kim2014flipping, qureshi2022hydra, you2019mrloc, son2017making, wang2021discreet, yaglikci2022hira}) aim \revcommon{to} mitigate RowHammer at low area overhead.\agycomment{find more to cite here} 
Unfortunately, to achieve low area overhead, these works 
cause {(prohibitively)} large performance overheads as DRAM chips become more vulnerable to RowHammer~\rowHammerGetsWorseCitations{}. Therefore, it is important to provide {a} scalable RowHammer solution whose area overhead and performance overhead remain low \revcommon{as} DRAM chips \revcommon{become more vulnerable to RowHammer}.}

\Copy{C1intro}{
\textbf{Our goal} is to prevent RowHammer bitflips at low performance, energy, and area overhead\revcommon{s} in modern and future DRAM-based systems with high RowHammer vulnerability. To this end, we propose a new low-cost and scalable counter-based RowHammer mitigation mechanism, \emph{\XLong{} (\X{})}. \X{} leverages \agy{our} \textbf{key observation} on benign workloads' \revi{and RowHammer attacks'} memory access patterns. Many workloads \changev{\ref{c:c1}}\revi{(both benign workloads and RowHammer attacks)} \agy{tend to} access DRAM rows with the \emph{same} row address in \emph{multiple} DRAM banks at \emph{around the same time} because i)~modern memory address mapping schemes interleave consecutive cache blocks across different banks \revcommon{(\secref{sec:dram_background})} and ii)~workloads tend to access cache blocks in close proximity around the same time due to the spatial locality in their memory accesses \revcommon{(\secref{sec:motivation})}.
}

Leveraging this observation, \X{}'s \textbf{key idea} is 
\agy{to {use} a \emph{{single shared} activation counter} {to track activations to} the rows with the same row \gf{ID} (i.e., same row address) in all DRAM banks}.
\agy{By doing so,} \X{} i)~retain\agy{s} the performance- and energy-efficiency benefits of \agy{counter-based RowHammer mitigation mechanisms}
(\secref{sec:evaluation}), and ii)~incur\agy{s} low area cost, as it requires \emph{only} a small number of counters to keep track of many aggressor rows (e.g., \param{2720} counters \agy{instead of \param{43520}~\cite{park2020graphene}} at a\agy{n} \gls{nrh} of \param{1000}). 

%

\textbf{Key Mechanism.}
At a high level, \agy{\X{} maps} DRAM rows {that have} the same \agy{row \gf{ID} in different banks} (which we call \emph{sibling rows}) \agy{to the same \X{} counter.}
In each \X{} counter, we store i)~the \emph{sibling activation vector} that contains as many bits as the number of \agy{banks}
(e.g., 16 bits if there are 16 banks), and ii)~the \emph{row activation count}. 
\agy{\X{} tracks {only} the maximum (i.e., the worst) activation count {of} the activation counts of sibling rows.} Before the activation count value reaches \gls{nrh}, \X{} \agy{preventively} refreshes \agy{all potential victim rows of each sibling row}
\agy{and thus prevents any} potential RowHammer bitflips.
While \X{} tracks the maximum activation count \hluo{of sibling rows, it also does} \emph{not} increment the \hluo{row activation count} unnecessarily with each sibling row activation. This way, \X{} \agy{reduces} the number of unnecessary \emph{preventive refresh} operations, \agy{lowering} its performance and energy overheads. \X{} is {completely} implemented inside the memory controller and {therefore} does \emph{not} require any modifications to existing DRAM chips {or software}.

\textbf{Key Results.}
We rigorously evaluate \X{}'s i) impact on system performance and energy consumption using cycle-accurate memory system simulations {(with Ramulator~\cite{kim2016ramulator,ramulatorgithub,luo2023ramulator,ramulator2github}),} executing a diverse set of \reviii{\param{62}} single-core and \reviii{\param{62}} 8-core multi-programmed workloads from SPEC CPU2006, SPEC CPU2017, \nb{TPC, MediaBench, and YCSB} benchmark suites and memory-intensive microbenchmarks, and ii) area overhead using \gf{CACTI~\cite{cacti}.} 
{We model \X{}'s hardware design (RTL) in Verilog and evaluate its circuit area and latency overheads using modern ASIC design tools.}
We compare \X{} to four state-of-the-art {RowHammer} mitigation mechanisms. 
We make \param{{four}} key observations from our evaluation. 
First, at a near-future RowHammer threshold of 1K, \X{} incurs \emph{only} 1)~\reviii{\param{0.58}}\% average (\reviii{\param{32.00}}\% maximum) performance and \reviii{\param{1.66}}\% average (\reviii{\param{2.02}$\times{}$} maximum) DRAM energy overheads across \reviii{\param{62}} single-core workloads, and 
2)~\param{0.77}\% average (\param{32.97\%} maximum) performance and \param{2.12}\% average (\param{2.17$\times{}$} maximum) DRAM energy overheads across \reviii{\param{62}} 8-core workload mixes compared to a system without any RowHammer protection, {while} requiring only \param{18.93} KiB of storage. 
Second, \X{} scales well into the future \agy{for DRAM chips with} extreme\agy{ly low} RowHammer thresholds: {e.g.,} at a RowHammer threshold of \emph{only} \param{125}, \X{}’s performance \agy{and energy} overheads \agy{are} \reviii{\param{1.45}}\% and \reviii{\param{1.27}}\%\agy{, respectively,} on average for single-core workloads, {while} requiring \param{151.41} KiB of storage. 
{Third, a}t {the} \gls{nrh} {of 125}, 
\agy{\X{} {performs very similarly to the best prior performance- and energy-efficient RowHammer mitigation mechanism while requiring \param{22.72}$\times{}$ smaller chip area.}} 
{Fourth,} \X{} scales well with the number of DRAM banks. At {the} \gls{nrh} {of 125}, \X{} incurs \reviii{\param{1.58}}\%, \reviii{\param{1.50}}\%, and \reviii{\param{2.60}}\% performance overheads for 16-, 32-, and 64-bank systems across all \reviii{\param{62}} single-core workloads, respectively.
{Our evaluation of \X{}'s circuit latency shows that \X{} could be implemented off the critical path in the memory controller. \X{}'s latency (\SI{1.22}{\nano\second}) is easily overlapped with the latency (\SI{2.5}{\nano\second}~\cite{jedec2017ddr4}) of issuing two successive DRAM row activation commands ($tRRD$).}
We open source our simulation infrastructure and all datasets {at \url{https://github.com/CMU-SAFARI/ABACuS}} to enable reproducibility and help future research.

This work makes the following key contributions:
\begin{itemize}
    \item We show that it is possible to leverage \agy{benign} workload access patterns to \agy{prevent RowHammer bitflips} at low \agy{overhead in terms of} performance, energy, and area, even for DRAM chips with very high RowHammer vulnerability.
    \item We develop \X{}, a new low-cost and scalable RowHammer mitigation mechanism. \X{} prevents RowHammer bitflips with {small average} performance and energy overheads {while} using \agy{significantly} small\agy{er} in-processor-chip storage \agy{compared to state-of-the-art RowHammer mitigation mechanisms} at very low RowHammer thresholds (i.e., 1K to 125).
    \item We evaluate the performance, energy, and area overheads of four state-of-the-art RowHammer mitigation mechanisms. We show that \X{} performs very similarly to the best-performing state-of-the-art mechanism at a {much} smaller {(e.g., 22.72$\times{}\!$) chip} area overhead.
    We model \X{}'s hardware design (RTL) in Verilog and evaluate its circuit area and latency overheads using modern ASIC design tools.
\end{itemize}
\glsresetall{}
\section{Background}
\label{sec:background}
\subsection{DRAM \agy{Organization and Operation}}
\label{sec:dram_background}

\begin{figure}[h]
    \centering
    \includegraphics[width=\linewidth]{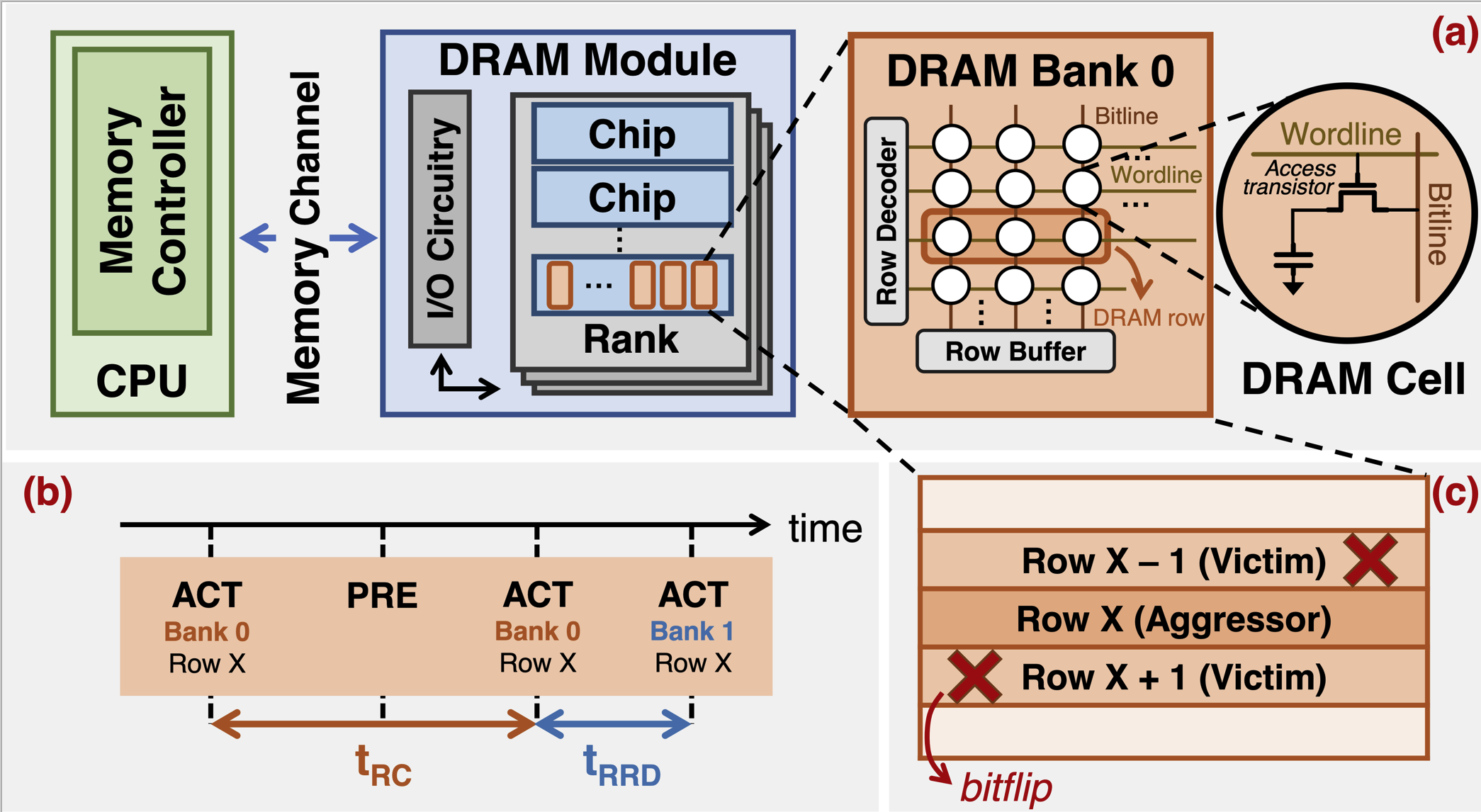}
    \caption{DRAM organization {(a), timing parameters (b), and RowHammer (c)}}
    \label{fig:dram_organization}
\end{figure}

\noindent
\textbf{Organization.}
\agy{\figref{fig:dram_organization}a shows the organization of DRAM-based memory systems. A memory channel connects the processor (CPU) to a set of DRAM chips.}
\yct{
\agy{This set of DRAM chips forms one or more} DRAM ranks that operate in lockstep.
Each chip has multiple DRAM banks, where DRAM cells are organized as a two-dimensional array of DRAM rows and columns. 
\agy{A DRAM cell is connected to the row buffer via a wire called bitline.}
The DRAM cell \agy{stores one bit of} data \agy{in the form of} electrical charge in \agy{a} capacitor. \agy{The access transistor, controlled by the wordline, connects the cell to the bitline.}
}

\noindent
\textbf{Operation.}
{
To serve main memory requests, the memory controller issues DRAM commands, e.g., row activation ($ACT$), bank precharge ($PRE$), data read ($RD$), data write ($WR$), and refresh ($REF$).}
\nb{To read or write data, the memory controller first issues an $ACT$ \gf{command alongside the bank address (i.e., bank ID) and row address (i.e., row ID) corresponding to the memory request's address, which opens the row.} 
When a row is activated, its data is copied to the row buffer. The memory controller can read/\gf{write}
data {at} cache block (512 bits) granularity from/to the row buffer using {a sequence of} $RD$/$WR$ commands. A subsequent access to the same row (i.e., a \emph{row hit}) can be served quickly without issuing another $ACT$ to the same row. To access another row (i.e., to serve a \emph{row conflict}), the memory controller must issue a precharge command and close the open row.} 

\noindent
\revcommon{\textbf{Periodic Refresh.}}
\yct{
\agy{DRAM cells are inherently leaky and thus lose the stored electrical charge over time. To} maintain data integrity, a DRAM cell \agy{is periodically refreshed} every {\gls{trefw}, which is typically} \SI{64}{\milli\second} (e.g.,~\cite{jedec2012ddr3, jedec2017ddr4, micron2014ddr4}) or \SI{32}{\milli\second} (e.g.,~\cite{jedec2015lpddr4, jedec2020ddr5, jedec2020lpddr5}). 
To \agy{timely} refresh all cells, the memory controller \agy{periodically} issues a refresh \agy{($REF$)} command every \gls{trefi}, \agy{which is typically} \SI{7.8}{\micro\second} \agy{(e.g.,~\cite{jedec2012ddr3, jedec2017ddr4, micron2014ddr4}) or} \SI{3.9}{\micro\second} \agy{(e.g.,~\cite{jedec2015lpddr4, jedec2020ddr5, jedec2020lpddr5})\gf{.} Upon receiving a $REF$ command, the DRAM chip internally refreshes {multiple} DRAM rows for a \gls{trfc} amount of time}. 
}

\noindent
\nb{\textbf{DRAM Timing Parameters.} 
The memory controller schedules DRAM commands according to certain timing parameters to guarantee correct operation~\cite{jedec2020ddr5,jedec2020lpddr5,jedec2015lpddr4, jedec2015hbm,jedecddr,jedec2017ddr4,jedec2012ddr3,salp,lee2013tiered,ipek2008self}. \agy{In addition to \gls{trefw}, \gls{trefi}, and \gls{trfc},} two \agy{other} timing parameters are relevant for this work: i)~\gls{trc} and ii)~\gls{trrd} (Fig.~\ref{fig:dram_organization}b).}

\noindent
\nb{\textbf{Bank-Level Parallelism.} Main memory accesses that target different banks \agy{can} proceed concurrently~\cite{salp,zhang2000permutation}. 
\atb{Modern} address mapping schemes {(e.g.,~\cite{hur2019adaptive,kaseridis2011minimalistic,liu2018get,ghasempour2016dream})} aim to interleave consequently addressed cache blocks {across} different banks to exploit \emph{bank-level parallelism}~\cite{mutlu2008parbs,salp}.} \ext{\figref{fig:address-mapping} depicts an address mapping that spreads consecutive cache blocks in main memory to DRAM rows with the same row ID in different banks.}

\subsection{RowHammer Mitigation {Techniques}}

{To prevent RowHammer bitflips and protect computing systems against RowHammer attacks, prior works propose \gf{different RowHammer mitigation} mechanisms~\mitigatingRowHammerAllCitations{}.} \agy{These works trigger their countermeasure (e.g., refreshing potential victim rows {or throttling accesses to potential aggressor rows}) based on either i)~the result of a probabilistic procedure or ii)~tracking \gf{the number of times DRAM rows are activated (i.e.,} row activation counts\gf{)}. While probabilistic procedures can be implemented at low chip area cost, they incur prohibitively large performance overheads when configured for sub-1K \gls{nrh} values~\cite{kim2020revisiting, yaglikci2022hira,yaglikci2021blockhammer}. Prior works propose several different row activation tracking mechanisms that detect the {frequently-activated} set of rows. Unfortunately, while providing better performance than the probabilistic mechanisms, the chip area overhead of these {row-activation-count-tracking} mechanisms significantly increases as DRAM chips become more vulnerable to RowHammer~\cite{qureshi2022hydra,mutlu2022fundamentally-DAC}.\footnote{\agy{Hydra~\cite{qureshi2022hydra} is an exception in this classification as it incurs a low {chip} area overhead while tracking row activation counts. Hydra achieves this by storing the counters in {the} DRAM array and caching them in the memory controller. \secref{sec:motivation} discusses Hydra's scalability limitations.}}}

\noindent
\textbf{Frequent Item Counting.}
\yct{A na\"{i}ve, area-inefficient tracking method to detect possible aggressor rows is to store the activation count of each DRAM row in \agy{a dedicated} counter. However, this method leads to impractical on-chip area overheads when used to protect modern, high-density DRAM modules. For example, 8-bit counters for a modern DDR4 rank with $2^{21}$ rows~\cite{jedec2017ddr4} would require 2 MiB on-chip storage and {a newer and denser DDR5 rank with $2^{23}$ rows~\cite{jedec2020ddr5} would require an even larger 8 MiB on-chip storage.} Fortunately, the problem of tracking \agy{the frequently activated DRAM} rows can be interpreted as a frequent item counting problem and {can be solved using} more area-efficient algorithms. For example, {the} Misra-Gries algorithm~\cite{misra1982finding} can be implemented in hardware to accurately track aggressor rows using \agy{a relatively small number of} counters to detect \agy{potential} aggressor rows, \agy{\omcri{and its variants are} adopted by several} prior RowHammer mitigation mechanisms~\agy{\cite{park2020graphene, kim2022mithril, saileshwar2022randomized, saxena2022aqua, marazzi2022protrr}}. 
}

\section{Motivation}
\label{sec:motivation}

{Repeatedly activating and precharging (hammering) a DRAM row \emph{at least} \gls{nrh} times in a refresh window induces one or multiple bitflips in that row.} As DRAM chips become more vulnerable to RowHammer {(i.e., the chip's rows have smaller \gls{nrh} values)}, fewer {hammers} can induce bitflips. {Even though the number of activate and precharge commands that the memory controller can issue in a refresh window remains the same,} more rows can be concurrently hammered {\gls{nrh} times at a smaller \gls{nrh}}. 
As RowHammer vulnerability increases, state-of-the-art {counter-based} RowHammer mitigation mechanisms need to track more DRAM rows and implement more activation counters. 

{A common method of increasing memory bandwidth and capacity is to increase the number of DRAM banks~\cite{salp, chang2014improving, mutlu2008parbs, subramanian2016bliss, subramanian2014bliss, lee2009improving}. {However, as the number of banks increases, counter-based mechanisms incur increasing chip area overhead (as the number of rows to track increases linearly with the number of banks).}}

\noindent
\textbf{Limitations of Prior Work.}
\agy{Several prior works~\cite{qureshi2022hydra,seyedzadeh2017cbt, seyedzadeh2018cbt, kim2014flipping, you2019mrloc, kang2020cattwo, devaux2021method} aim \gf{to mitigate} RowHammer at low area overhead by implementing {a limited set of row activation} counters {(i.e., fewer counters than there are rows in the DRAM module)} at the cost of reduced tracking accuracy. {However, a RowHammer mitigation countermeasure (e.g., preventively refreshing potential victim rows or throttling accesses to potential aggressor rows) fundamentally consumes memory bandwidth (e.g., by making the DRAM module unavailable for memory demand requests while performing refresh or by throttling memory demand requests) and reduced tracking accuracy exacerbates the number of countermeasures deployed by the mitigation mechanism. Thus,} these mechanisms occupy a significant portion of main memory bandwidth and thus incur large system performance and DRAM energy overheads {as} they are configured for small \gls{nrh} values. To provide more insight into this problem, we describe the key drawback of one such state-of-the-art mechanism, Hydra~\cite{qureshi2022hydra}, as a concrete example.}
Hydra~\cite{qureshi2022hydra} maintains the activation count of each DRAM row in a physical location in main memory (i.e., in the DRAM chip{s}). 
To minimize the overheads of \agy{fetching the counters from the main memory\gf{,} Hydra implements a filtering and caching logic. The filtering logic groups a number of (e.g., 125) DRAM rows into row groups and assigns a counter to each row group called the \emph{group counter}.}
DRAM row activations update \emph{{only}} the {corresponding} group counters at the beginning of a refresh window. When a group counter exceeds a predetermined {\emph{group count threshold}} (e.g., \param{400}), the group counter's value is copied to the activation counters of rows in that group, {such that Hydra can track each row's activation count individually and deploy its countermeasure (preventively refreshing victim rows) more accurately (e.g., instead of preventively refreshing \emph{all} 125 rows in a group, Hydra can refresh one or several DRAM rows that are activated frequently in the group of 125 rows, depending on workload memory access patterns),} and the group counter is no longer queried. 

\gf{Hydra's} mechanism has \agy{two} \textit{key drawback\agy{s}}. 
\agy{First, Hydra} overestimates the activation count{s} of DRAM rows\gf{, causing} 
a {large} number of unnecessary refresh operations. {According to our system-level simulations~(in~\secref{sec:evaluation}), \emph{approximately half of} Hydra's preventive refresh operations are unnecessary for 62 single-core workloads at a very low RowHammer threshold of 125}. 
{Hydra overestimates activation counts of DRAM rows because m}odern memory-intensive workloads can rapidly increase the group counter value to the {group count} threshold value with \emph{only} a few activations to each DRAM row in the group. \revcommon{Such workloads can} cause the activation counters to overestimate the actual activation count of each row in the group by up to \param{396} (in Hydra's default configuration for an \gls{nrh} of $1K$). 
Therefore, Hydra often mistakenly refreshes the neighbors of {many} rows that will \emph{not} be activated as many as \gls{nrh} times. 
Second, suppose the counter of an accessed row is not cached in the memory controller. In that case, Hydra needs to fetch the counter from the main memory, which incurs additional memory latency for writing back the evicted counter and fetching the new counter. 
\agy{Both of these drawbacks} incur \agy{significant} performance and energy overheads {as Hydra increases the memory latency (i.e., the time it takes to serve a memory demand request) by \param{23.67\%} on average at \gls{nrh} = 125} {(as we show in detail in~\secref{sec:evaluation})}. 

\noindent
\subsection{{Motivational Analysis for \X{}}}
\Copy{C2/1}{\revii{We investigate memory access patterns of modern memory-intensive workloads \revi{and existing RowHammer attacks}{. We} observe that they activate DRAM rows with the same row address in multiple DRAM banks ({i.e.,} sibling rows) at around the same time. This observation motivates us to design a performance- and energy-efficient DRAM row activation count tracking mechanism at low area cost by implementing \emph{one shared activation counter} for all sibling rows. Implementing one shared activation counter reduces the number of counters required to track aggressor rows (and thus the area cost) by a factor of the number of banks (e.g., 16 in DDR4~\cite{jedec2017ddr4}) compared to the aggressor row tracking mechanisms used by the state-of-the-art performance- and energy-efficient RowHammer mitigations (e.g., Graphene~\cite{park2020graphene}{, Panopticon~\cite{bennett2021panopticon}, and PRHT~\cite{kim2023ddr5}}).} \revii{However, the shared counter may not accurately represent the activation counts of multiple sibling rows because the shared counter can store only one activation count. Misrepresenting the activation counts of sibling rows may induce performance and energy overheads due to unnecessary victim row refresh operations. In the remainder of this section, we show that 1) sibling rows are activated at around the same time, and 2) a shared activation counter can accurately represent the activation counts of multiple sibling rows.}}

\begin{figure*}[t]
    \centering
    \Copy{C1fig}{
    \includegraphics[width=0.8\textwidth]{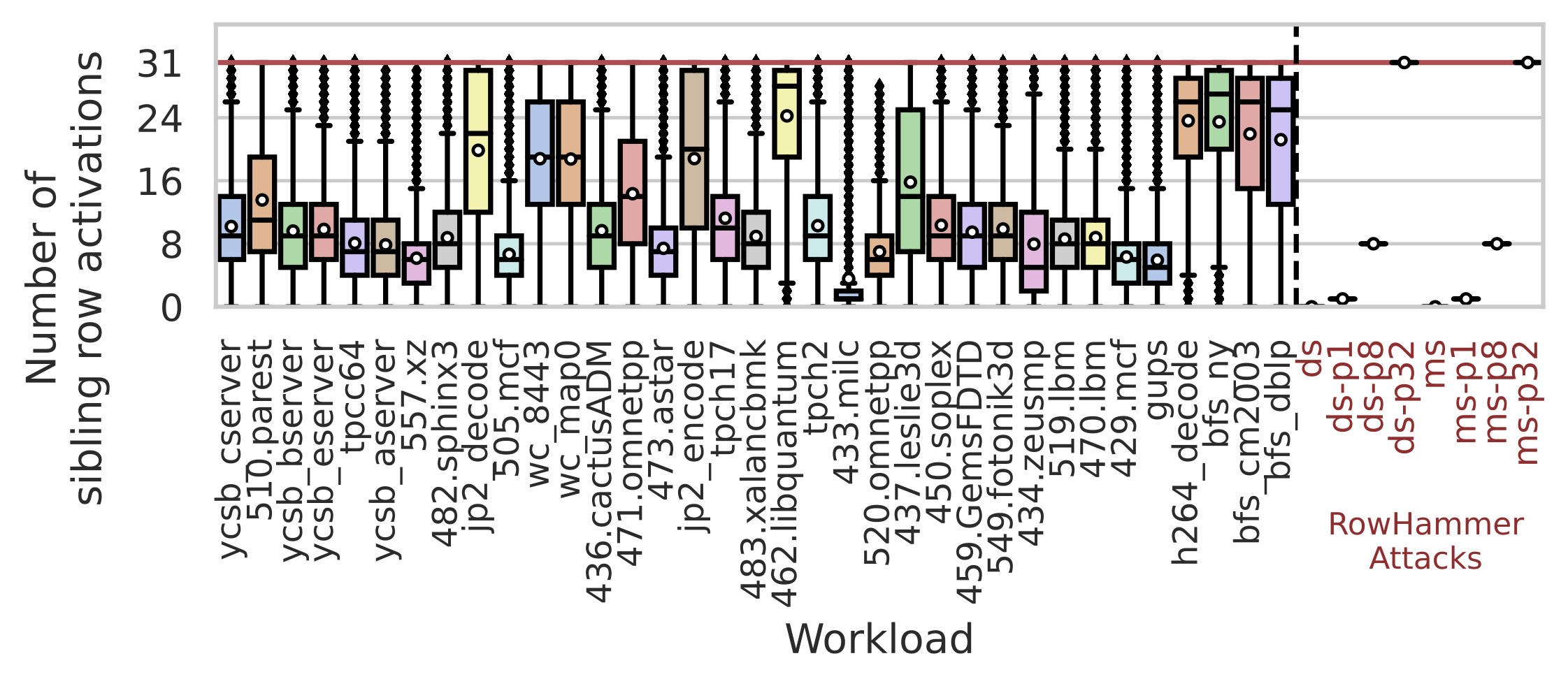}
    \caption{\revii{Number of sibling row{s activated before one sibling row is activated again}, averaged across all DRAM row activations for each simulated workload {and RowHammer attack} (x-axis).}}
    \label{fig:average_counter_values_single}
    }
\end{figure*}

We simulate \revii{34} memory-intensive workloads{, each having} more than two row buffer misses per kilo instructions (RBMPKI), \changev{\ref{c:c1}}\revi{and three variations of the double-sided (ds)~\cite{kim2020revisiting,orosa2021deeper,seaborn2015exploiting,pessl2016drama,razavi2016flip,jang2017sgx,tatar2018defeating} and many-sided (ms)~\cite{deridder2021smash,hassan2021utrr,frigo2020trrespass,cohen2022hammerscope,jattke2022blacksmith,kogler2022half} on a 32-bank system using the simulation methodology that we explain in \secref{sec:evaluation:methodology}. \revi{We carefully create memory traces (load and store requests that arrive at main memory) of double- and many-sided attacks 1) without prefetching, 2) with a simple prefetcher that prefetches the \emph{next cache line} (p1)~\cite{jouppi1990improving,smith1982cache}, 3) the \emph{next eight cache lines} (p8), 4) and the \emph{next 32 cache lines} (p32) for every load request.}
{We name a RowHammer attack trace as the concatenation of its type (ds or ms) and the prefetcher configuration used in creating the trace (p1, p8, or p32). For example, \emph{ds-p32} is the double-sided RowHammer attack with the next 32 cache line prefetcher.}
\changev{\ref{c:c2}}\revii{\figref{fig:average_counter_values_single} shows how many sibling rows get activated before one of the sibling rows is activated \emph{{again}}, on average across all DRAM row activations (y-axis) for each simulated workload (x-axis).\footnote{\label{fn:boxplot}{{{The box is lower-bounded by the first quartile (i.e., the median of the first half of the ordered set of data points) and upper-bounded by the third quartile (i.e., the median of the second half of the ordered set of data points).
The \gls{iqr} is the distance between the first and third quartiles (i.e., box size).
Whiskers show the central 90th percentile of the distribution.}}}} {Benign workloads are ordered from left to right in increasing memory intensity (in terms of row buffer misses per kilo instructions) in the figure.} We highlight the highest possible y-axis value (31) with a red line on the plot. A workload with a bar closer to this line indicates that the workload accesses \emph{all} sibling rows at around the same time.}}

\Copy{C2/3}{
\revii{In the upper extreme case (at $y=31$)\footnote{{Because a DRAM row has 31 sibling rows in a system with 32 banks, the y-axis value cannot exceed 31.}}, the workload \emph{always} activates \emph{all} sibling rows once before activating any sibling rows for the second time. This property of the workload makes it a good fit for using a single activation counter shared between sibling rows. To prevent RowHammer bitflips in victim rows (i.e., to know when an aggressor sibling row has been activated \gls{nrh} times), the shared counter stores the highest activation count across all sibling rows. Because the workload activates all sibling rows once before activating any other for the second time, the shared counter accurately represents the activation count of every sibling row (as the difference between the minimum and the maximum activation count across sibling rows is 1). In the lower extreme case (at $y=0$), the workload \emph{never} activates two or more sibling rows. For this type of a workload, the single shared activation counter's value 
misrepresents almost all of the sibling rows' activation counts (which are 0).}}

\ignore{
\agy{To \atb{prevent} RowHammer bitflips at low area, performance, and energy overheads, \revii{our idea is to share a DRAM row activation counter across DRAM rows with the same row address in multiple DRAM banks (sibling rows). We identify two requirements for the shared activation counter to 1) securely and 2) performance- and energy-efficiently prevent RowHammer bitflips.}\ignore{the row activation patterns are similar across DRAM banks.}
\revii{First, to take timely action in securely preventing RowHammer bitflips, this shared counter must accurately track the sibling row with the highest activation count. This ensures that all potential victim rows are preventively refreshed when the shared activation counter reaches the RowHammer threshold (\gls{nrh}). However, because a single activation counter is shared across sibling rows .}
\atbcomment{We need to introduce preventive refreshes in the background.}
\revii{Second, to , the activation counts of other sibling rows must be close to the highest activation count.}
To \revii{understand how accurately the aggressor row's activation count can indicate the activation counts of its siblings}, we simulate \revii{34} memory-intensive workloads with more than two row buffer misses per kilo instructions (RBMPKI) on a 32-bank system using the simulation methodology that we explain in \secref{sec:evaluation:methodology}.}
\ignore{
\agy{\figref{fig:average_counter_values_single} shows the number of banks accessed with the same row \gf{ID} in a window of 128 row activations (y-axis) for each workload (sorted in increasing RBMPKI order on the x-axis) in a box plot.\footnote{\label{fn:boxplot}{{The box is lower-bounded by the first quartile (i.e., the median of the first half of the ordered set of data points) and upper-bounded by the third quartile (i.e., the median of the second half of the ordered set of data points).
The \gls{iqr} is the distance between the first and third quartiles (i.e., box size).
Whiskers show the central 90th percentile of the distribution.}}}}

\begin{figure*}[!t]
    \centering
    \includegraphics[width=0.8\textwidth]{figures/bank-cnts-med-128.pdf}
    \caption{Number of banks that a workload accesses with the same row \gf{ID} in a window of 128 row activations.}
    \label{fig:bank_cnts}
\end{figure*}
}
}

\begin{figure*}[!t]
    \centering
    \Copy{C2fig}{
    \includegraphics[width=0.8\textwidth]{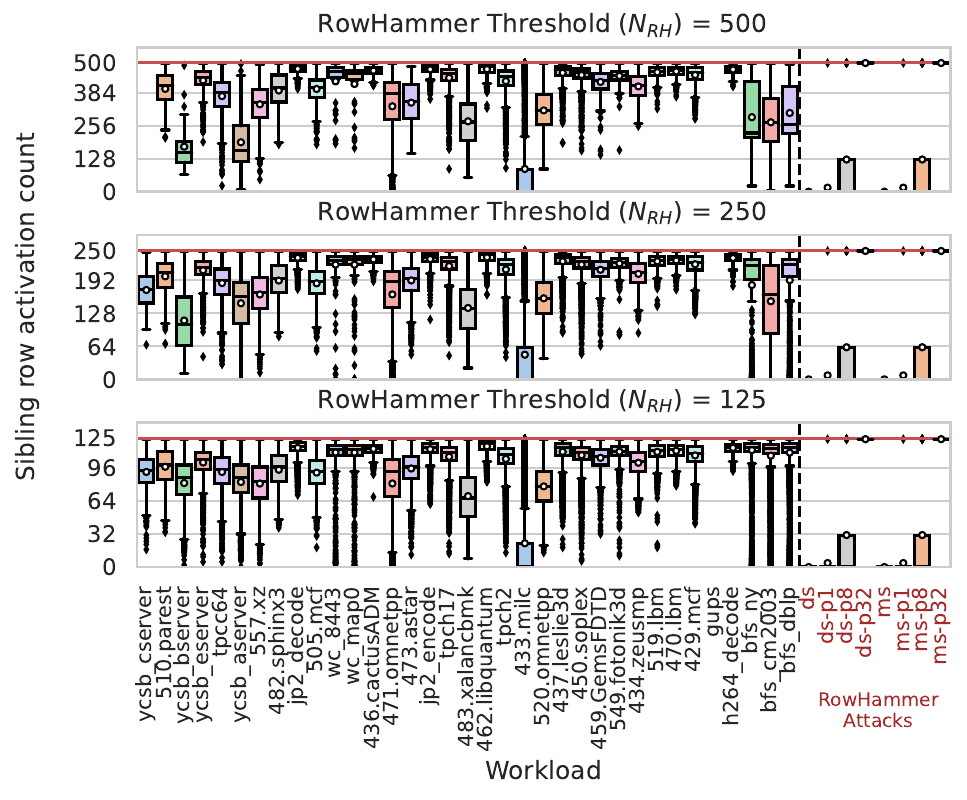}
    \caption{\revii{The distribution of the number of activations a sibling row receives before any sibling row gets activated \gls{nrh} times for three different \gls{nrh} values for each {benign} workload {and RowHammer attack} (x-axis).}}
    \label{fig:distr_counter_values}
    }
\end{figure*}

\Copy{C2/4}{
\revii{We make \revi{three} key observations from \figref{fig:average_counter_values_single}. First, on average across all workloads, \param{12.8} sibling rows get activated before any sibling receives {another} activation. We observe that {some} workload{s} activate at least \param{\emph{three}} sibling rows \emph{once}, while some activate up to \emph{\param{25}} sibling rows (out of 31), before activating any sibling row \emph{{again}}. Second, the average sibling row activation count does \emph{not} significantly correlate with the memory intensity of the workload.} \revi{Third, the RowHammer attacks can activate up to 31 sibling rows before activating any sibling row {again} due to the prefetch requests generated by the simple next cache line prefetcher.} \agy{From these observations, we conclude that after accessing a row with the address $R$ in a bank, the rows at address $R$ in other banks (i.e., sibling rows) are also \gf{likely to be} activated.} \revii{We hypothesize that} this access pattern occurs due to two properties: i)~the memory address mapping schemes that \revii{aim to increase} {memory} bank-level parallelism to improve system performance (\secref{sec:dram_background})\atbcomment{should we describe in more detail?} and ii)~the \agy{intrinsic spatial locality in workloads' {main} memory accesses}. 

\revii{To strengthen our motivation for sharing an activation counter between sibling rows, we plot the distribution of the activation count of each sibling row when at least one sibling row gets activated \gls{nrh} times. {In other words, one point in the distribution is an activation count of a row. One of this row's siblings has been activated \gls{nrh} times.} \figref{fig:distr_counter_values} shows how many times a sibling row gets activated (y-axis) before one of the sibling rows gets activated \gls{nrh} {times for \gls{nrh} =} 500, 250, and 125 (different subplots) {across benign workloads and RowHammer attacks} (x-axis). We highlight the highest possible y-axis values for each \gls{nrh} value.}
}

\Copy{C2/5}{
\revii{We make \param{two} observations from \figref{fig:distr_counter_values}.\changev{\ref{c:c3}} First, a sibling row gets activated \param{99}, \param{194}, and \param{370} times for \gls{nrh} values of 125, 250, and 500, on average across all workloads. This indicates that when any sibling row gets activated \gls{nrh} times, the activation counts of all sibling rows are {(very)} close to \gls{nrh}. Second, as \gls{nrh} reduces, the gap between the average activation count of a sibling row and the sibling row with the highest activation count becomes smaller in proportion. For example, the \texttt{bfs\_cm203} workload, on average, activates sibling rows \param{272} (54.4\% of the \gls{nrh} of 500) and \param{108} (86.4\% of the \gls{nrh} of 125) times for \gls{nrh} values of 500 and 125, respectively.}\footnote{\Copy{C3f1}{\reviii{The gups workload \emph{cannot} activate any same row 125 times because a workload is limited in the number of DRAM row activations it can issue to a DRAM chip by DRAM timing parameters (e.g., four row activation window, $t_{FAW}$~\cite{jedec2017ddr4}) and gups' row activations are randomly and evenly distributed to all 128K DRAM rows. A workload that fully exercises the available DRAM row activation bandwidth can \emph{only} issue \param{12'190'476} activate commands in a refresh window (\SI{64}{\milli\second}) due to $t_{FAW}$ (\SI{21}{\nano\second}) timing constraint~\cite{jedec2017ddr4}. Therefore, randomly and evenly distributing \param{12'190'476} activate commands to 128K rows would activate a row at most \param{94} times.}}}

\revii{From our analysis, we conclude that a single shared activation counter, which stores the highest activation count among the activation counts of all sibling rows, can {reasonably} accurately represent the activation count of all sibling rows. This property of the shared activation counter becomes stronger as \gls{nrh} reduces from 500 to 125.}
}

\ignore{
\agy{From \figref{fig:average_counter_values_single}, we make two observations. First, most of the workloads (all except 557.xz and 520.omnetpp) access more than one bank with the same row \gf{ID} on average. Second, the workloads that exhibit larger memory intensity (in terms of RBMPKI) tend to access more banks using the same row \gf{ID}. For example, the right-most three workloads access the same row \gf{ID} in all 32 banks in a window of 128 row activations.}
This access pattern \agy{occurs because of two properties:} 
i)~the memory address mapping schemes that target increasing the bank-level parallelism to improve system performance (\secref{sec:dram_background}) and ii)~the \agy{intrinsic spatial locality in the workload's memory accesses}.
\agy{From these observations, we conclude that after accessing a row with the address $R$ in a bank, the rows at address $R$ in other banks are also \gf{likely to be} activated.}
}
\noindent

\section{Mechanism}
\label{sec:mechanism}

\noindent
\textbf{Overview.}
\X{} is designed to prevent RowHammer bitflips at \hluo{low performance, energy, and area overhead}. Achieving low performance and energy overheads requires accurately identifying aggressor rows and preventively refreshing victim rows \emph{only} when necessary. To this end, \X{} adopts the \hluo{Misra-Gries algorithm~\cite{misra1982finding} (\secref{sec:background}) to track aggressor rows, similar to} prior work~\cite{park2020graphene, kim2022mithril, saileshwar2022randomized, saxena2022aqua, marazzi2022protrr}. However, Misra-Gries alone \emph{cannot} prevent RowHammer bitflips at low area cost (\secref{sec:evaluation-area}).
Thus, \X{} {performs} Misra-Gries tracking {using} \emph{shared activation counters} \hluo{to significantly reduce the area overhead of implementing a large number of counters.} 

\begin{figure}[ht]
    \centering
    \includegraphics[width=\linewidth]{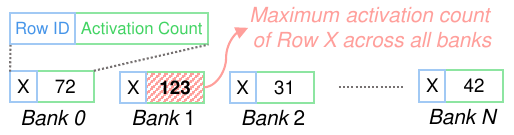}
    \caption{Maximum (worst) activation count of sibling rows}
    \label{fig:maximum_activation_count}
\end{figure}

\X{}'s key idea is to share a 
\hluo{\emph{row activation counter}} {among} rows that have the same \hluo{row ID} across 
all banks (which we call \emph{sibling rows}). \hluo{The shared row activation counter tracks the maximum activation count among the sibling rows, as illustrated in~\figref{fig:maximum_activation_count}. \X{} preventively refreshes all the 
neighboring rows of {all} the sibling rows tracked by the same row activation counter \omcri{just enough} \emph{before} the row activation counter's value reaches \gls{nrh} within a \gls{trefw} to prevent RowHammer bitflips (i.e., none of the sibling rows' activation count reaches \gls{nrh} within \gls{trefw}).}

While \X{} tracks the maximum activation count \hluo{{among} sibling rows, it does} \emph{not} unnecessarily increment the \hluo{shared row activation counter} with each sibling row activation. For example, when a workload activates multiple sibling rows \emph{only once} (which is a common behavior we observe in~\secref{sec:motivation}), it is sufficient for \X{} to increment the shared activation counter \emph{by only one}. After a sibling row is activated and \X{} increments the \hluo{shared row activation counter}, other sibling rows can be activated \emph{at most once} before the shared counter is incremented again. To allow for multiple sibling rows to be activated \emph{without} unnecessarily incrementing the \hluo{shared row activation counter}, \X{} \hluo{maintains a bit vector for the counter, \emph{sibling activation vector}, that stores} which sibling rows were activated \emph{since the shared counter was last incremented}. \X{} increments the \hluo{shared row activation counter} only when the bit corresponding to the activated sibling row is already set (i.e., the sibling row was activated once since the shared counter was last incremented).

\subsection{{\X{}'s Hardware Design}}

\begin{figure}[h]
    \centering
    \includegraphics[width=\linewidth]{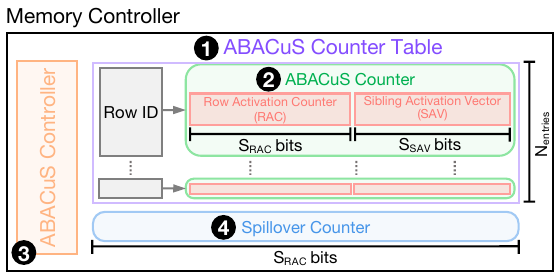}
    \caption{Key components of \X{}}
    \label{fig:abacus_key_components}
\end{figure}

\figref{fig:abacus_key_components} presents \X{}'s key components. \X{} is placed inside the memory controller. The \X{} counter table contains (\dingOne{}) multiple \X{} counters{, each mapped to a row ID}. {There are exactly $N_{entries}$ \X{} counters in the \X{} counter table.} An \X{} counter (\dingTwo{}) consists of a row activation counter (RAC) {of size $S_{RAC}$ bits} and a sibling activation {(bit)} vector (SAV) {of size $S_{SAV}$ bits}. 
The \X{} controller (\dingThree{}) dynamically maps \omcri{(not shown in the figure)} a {row ID} to a counter during runtime and uses a \emph{spillover counter} (\dingFour{}) to track the maximum activation count of all DRAM rows that do \emph{not} have an \X{} counter assigned (\secref{sec:mechanism-operation}). We explain how we determine the sizes of each key component ($S_{RAC}$, $S_{SAV}$, and $N_{entries}$) in~\secref{sec:mechanism-configuring}. 

The row activation counter (RAC) in an \X{} counter (\dingTwo{}) stores the maximum activation count across all sibling rows' activation counts. The sibling activation vector (SAV) stores the sibling activation bits used by \X{} to increment the RAC {\emph{only} when necessary (see~\secref{sec:mechanism})}. The \X{} controller updates RAC and SAV to ensure the maximum activation count {among} all sibling rows \revcommon{is tracked in the RAC}.

\subsection{Operation {of \X{}}}
\label{sec:mechanism-operation}

\hluo{We} describe \X{}'s operation in \param{five} key steps: i) initialization, ii) \X{} counter table search, iii) \X{} counter update, iv) \X{} counter replacement, and v) periodic reset. 

\noindent
\textbf{(1) Initialization {and Reset}.} Initially {(at system power on) and after periodic \X{} counter table reset (Step 5)}, no DRAM row {is} activated for the last \gls{trefw}. Thus, row activation counters (RACs) and sibling activation vectors (SAVs) in all \X{} counters and the spillover counter {all} store 0. 

\begin{figure*}[!th]
    \centering
    \includegraphics[width=\textwidth]{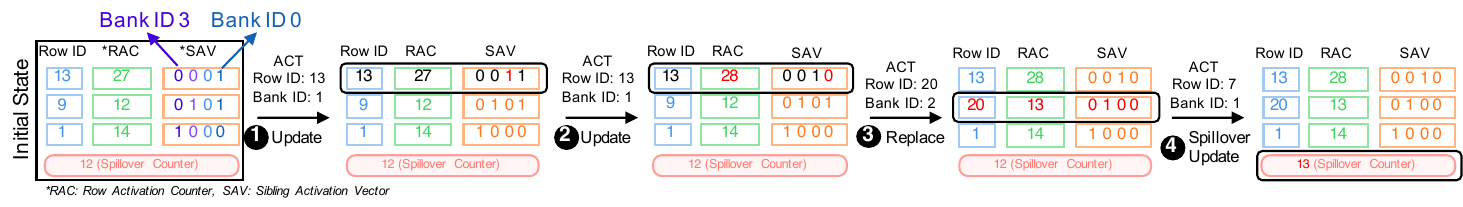}
    \caption{\X{} workflow using four \gls{act} commands. We highlight state changes using black boxes and red text.}
    \label{fig:abacus-example}
\end{figure*}

\noindent
\textbf{(2) \X{} Counter Table Search.} The memory controller issues an \gls{act} command to a row \gf{ID} in a bank. Consequently, the \X{} controller searches all row \gf{ID} mappings to find if the activated row is already tracked by an \X{} counter. {No row is tracked immediately after initialization and reset. Thus, the \X{} controller needs to map a row ID to an \X{} counter. To find the counter that will be mapped to the activated row's ID, the \X{} controller looks for an \X{} counter whose RAC stores the same value as the spillover counter's value. If a RAC and the spillover counter have the same value, the RAC's row ID is replaced with the activated row{'s ID} (Step 3). In contrast, if an \X{} counter already tracks the row ID, the \X{} controller updates the matching counter (Step 4). In case no RAC value equals the spillover counter's value (i.e., the activated row's ID \emph{cannot} be mapped to an \X{} counter) and there is no \X{} counter that already tracks the activated row's ID, the \X{} controller increments the spillover counter value.}
\yct{When the spillover counter reaches a predefined value of {the} refresh cycle threshold {(RCT)}, \X{} issues \gls{trefw}/\gls{trefi} \revcommon{\gls{ref}} commands to refresh all DRAM rows {in the DRAM rank} and resets all counters. \revcommon{We call the time when the memory controller refreshes all DRAM rows due to the spillover counter reaching the RCT a \emph{refresh cycle}.}}

\noindent
\textbf{(3) \X{} Counter {Mapping and} Replacement.} The \X{} controller maps the newly-activated row \gf{ID} to the matching {\X{} counter}. To correctly track the maximum activation count,
the \X{} controller i) initializes RAC with spillover counter value + 1, and ii) sets the bit in the SAV that corresponds to the bank \gf{ID} of the activated row.


\noindent
\textbf{(4) \X{} Counter Update.} The \X{} controller checks the SAV bit value corresponding to the activated row's bank ID. If the SAV bit is \emph{not} set (i.e., stores logic-0), {i.e., the activated row is activated \emph{for the first time} since the RAC was last incremented,} the \X{} controller sets the SAV bit.
If the SAV bit is set (i.e., stores logic-1), {i.e., the activated row is activated \emph{again} since the RAC was last incremented,} the \X{} controller increments the RAC by 1. 
{After incrementing the RAC, the \X{} controller sets} the SAV bit corresponding to the bank \gf{ID} {of the activated row} and {resets} all other bits. {This way, a set bit corresponding to the bank ID of the activated row in the SAV \emph{still} indicates that the corresponding row was activated \emph{once} since the RAC was last incremented.} If the RAC's value is \atb{a multiple of} the \emph{preventive refresh threshold (PRT)}, {i.e., one of the sibling rows tracked by the RAC was activated PRT times since all the sibling rows' victims were preventively refreshed,} \X{} performs preventive refresh operations to all \hluo{the victim rows {(the neighbors of the activated row)} in all banks}.


\noindent
\textbf{(5) Reset Period.} It is sufficient for \X{} to track the activation counts within a \gls{trefw}, after which all DRAM rows are refreshed. Therefore, we reset the \X{} counters and the spillover counter after every \gls{trefw} (i.e., \X{}'s reset period is \gls{trefw}). After periodic reset, \X{}'s state becomes as described in Step 1. 

\noindent
\textbf{Determining the Preventive Refresh Threshold (PRT).} \X{}'s preventive refresh threshold is set accordingly to prevent aggressor rows from being activated $N_{RH}/2$ times in a reset period. 
\revcommon{\X{} is configured in this way because} \X{} does \emph{not} \revcommon{precisely} know when each row is \revcommon{periodically} refreshed: A potential aggressor row might be \revcommon{hammered for $PRT-1$ times} before its neighbors are refreshed and \X{} is reset. After \revcommon{\X{} is reset}, an attacker can hammer \revcommon{the same aggressor row} for $2*PRT$ times, accumulating a total activation count of $2*PRT-1$ on the aggressor row. Thus, we set PRT to $N_{RH}/2$.

\subsection{Configuring \X{}}
\label{sec:mechanism-configuring}

Depending on the \hluo{system's} RowHammer vulnerability (\revcommon{typically} measured using \gls{nrh})\yct{,} \X{} has \param{three} key parameters that are configured at \hluo{design time}. First, the number of entries ($N_{entries}$) in the \X{} counter table. Second, the size of each row activation counter ($S_{RAC}$). Third, the size of each sibling activation vector ($S_{SAV}$).


\noindent
\textbf{Configuring the Number of Entries ($N_{entries}$).} We determine the number of entries based on how many rows can be hammered in one DRAM bank during one \X{} reset period (\revcommon{\SI{64}{\milli\second}}) given i) \revcommon{the preventive refresh threshold ($PRT$)}, ii) \gls{trc}, and iii) \gls{trfc} as described in a prior work that adopts Misra-Gries tracking~\cite{park2020graphene}. We first calculate how many \gls{act} commands can be issued by the memory controller in a reset period when the bank is \emph{not} unavailable due to periodic refresh ($N_{ACT}$) as $t_{REFW}*(1-t_{RFC}/t_{REFI})/t_{RC}$. Thus, at most $N_{ACT}/PRT$ rows can \revcommon{be activated PRT times} in \revcommon{an \X{}} reset period in a bank. Setting the number of entries to $N_{ACT}/PRT$ is sufficient to ensure that all hammered rows in an \X{} reset period in one bank are tracked in one of the counters. 

\noindent
\textbf{Configuring the Size of Row Activation Counters ($S_{RAC}$).} A row activation counter stores activation count values up to the number of activations that the memory controller can issue in a \gls{trefw}. Thus, a row activation counter should be $S_{RAC}=\lceil{}log_2((N_{ACT}*t_{RC})/t_{RRD})\rceil{}$ bits large. However, we can reduce $S_{RAC}$ to $\lceil{}log_2(PRT)\rceil{}$ given that we add an \emph{overflow} bit~\cite{park2020graphene} to each \X{} counter. Using the overflow bit, we make sure that \X{} preventively refreshes rows according to the Misra-Gries tracking algorithm~\cite{park2020graphene}: When a RAC reaches $PRT$, \X{} sets the overflow bit of the RAC and resets the RAC's value. The set overflow bit in a RAC indicates that \revcommon{\X{} should \emph{not} replace the row ID tracked by this RAC, even if the RAC's value equals the value of the spillover counter}. The overflow bit is reset after each reset period.

\noindent
\textbf{Configuring the size of sibling activation vectors ($S_{SAV}$).} A SAV contains as many bits as there are banks. Thus $S_{SAV}=N_{banks}$ bits. For example, there are 32 banks in a dual-rank DDR4-based system~\cite{jedec2017ddr4}, making $S_{SAV}$ 32-bit large.

\subsection{Example \X{} Workflow}

\figref{fig:abacus-example} shows an example \X{} workflow with three \X{} counters and four sibling rows. We show how four $ACT$ commands 
cause state changes in the three \X{} counters. The first $ACT$  command (\dingOne{}) updates the sibling activation vector (SAV) of its \X{} counter as the bank \gf{ID} corresponds to a zero bit in the SAV. The second $ACT$  command (\dingTwo{}) increments the row activation counter of its \X{} counter as the bank \gf{ID} corresponds to a non-zero bit in the SAV. The third $ACT$  command (\dingThree{}) replaces the second \X{} counter with row \gf{ID} 20 because i) row \gf{ID} 20 is not tracked by any counter, and ii) the spillover counter value is equal to the second \X{} counter's RAC value. The fourth $ACT$  command (\dingFour{}) increments the spillover counter as i) row \gf{ID} 7 is not tracked by any \X{} counter, and ii) the spillover counter value is smaller than {all} RAC values.

\section{Security Analysis}
\label{sec:security-analysis}

\X{} preventively refreshes the victim rows of a potential aggressor row before the aggressor row is activated \gls{nrh} times in a \gls{trefw}. Assuming \X{} {accurately} stores the maximum activation count across all sibling rows, the Misra-Gries-based tracking technique guarantees that no aggressor row is activated more than the preventive refresh threshold in a \gls{trefw}~\cite{park2020graphene}. \revcommon{\X{} {accurately} maintains the maximum activation count in the row activation counters because the row activation counter's value is incremented 1) when \emph{any} sibling row is activated for the first time, and 2) when a sibling row whose sibling activation vector bit is set (i.e., the sibling row was activated after the row activation counter's value was last incremented) is activated. Appendix~\ref{appendix:security-proof} formally analyzes and shows that the row activation counter always stores the maximum activation count.}

\section{{Accounting for RowHammer Blast Radius}}
\label{sec:blast}
\yct{An aggressor row can cause bitflips in victim rows that are \emph{not} physically adjacent~\cite{kim2014flipping, kim2020revisiting}. The impact of RowHammer on a victim row decreases \agy{and eventually disappears as the} physical distance \agy{between the victim and} the aggressor row\agy{s increases.}} 
\agy{To account for this characteristic, prior works define} \emph{blast radius} as the distance between an aggressor row and its furthest victim row~\agy{\cite{devaux2021method, yaglikci2021security, yaglikci2021blockhammer, kogler2022half, park2020graphene, kim2020revisiting, kim2014flipping, cojocar2020rowhammer, frigo2020trrespass, hassan2021utrr, jattke2022blacksmith, deridder2021smash, orosa2021deeper, hassan2022case, patel2022case}}.\agycomment{cite all papers that talk about blast radius. I added a few. Please double-check that they talk about blast radius (even if it is called something else) and find more.} \agy{A recent RowHammer attack,} Half Double~\cite{kogler2022half}, exploits blast radius to induce bitflips with a significantly lower activation count. 
\agy{To account for blast radius and address Half Double, \X{} 1)~preventively refreshes all potential victim rows within the blast radius and 2)~\agy{counts each preventive refresh as an additional activation}. \revcommon{We configure \X{} and other state-of-the-art mechanisms with a blast radius of one in our performance and energy evaluation (\secref{sec:evaluation:methodology})}.
\ext{\secref{sec:eval-blast-radius} in the Appendix evaluates \X{} and four other state-of-the-art mechanisms' performance overheads with blast radius up to 8.}
}

\section{\ext{Accounting for RowPress}}
\label{sec:rowpress}

\ext{RowPress~\cite{luo2023rowpress} is a recently discovered read disturbance phenomenon that affects many real DRAM chips. RowPress manifests as bitflips in DRAM rows when aggressor DRAM rows are kept open (i.e., activated) for a long time. RowPress can effectively reduce the RowHammer threshold of DRAM chips. For example, if a row an aggressor row is kept open for \SI{70.2}{\micro\second} each time it is opened for access, the number of activations needed to induce a bitflip is \emph{only} 100. Fortunately, RowPress can be mitigated with a combination of i) a secure RowHammer defense mechanism and ii) a memory controller that restricts how long a DRAM row can remain open~\cite{luo2023rowpress}. Thus, it is sufficient to configure \X{} for the correct \gls{nrh} value, considering the \gls{nrh}-reducing effect of RowPress and the maximum amount of time a row is allowed to remain open.}

\section{Hardware Implementation}
\label{sec:implementation}

\begin{savenotes}
\begin{table*}[!th]

    \caption{Area, energy, power of \X{} vs. state-of-the-art RowHammer mitigation mechanisms for {a 2-rank memory system}}
    \centering
    \scriptsize
    \renewcommand{\arraystretch}{0.92} %
    \resizebox{\linewidth}{!}{
    \begin{tabular}{l|ccccccc|ccccccc}
        \multirow{3}{*}{\textbf{Mitigation Mechanism}} & \multicolumn{7}{c}{\textbf{\gls{nrh} = 1000}} & \multicolumn{7}{c}{\textbf{\gls{nrh} = 125}} \\
            & \textit{SRAM} & \textit{CAM} & \multicolumn{3}{c}{\textit{Area}} & Access Energy & Static Power
            & \textit{SRAM} & \textit{CAM} & \multicolumn{3}{c}{\textit{Area}} & Access Energy & Static Power\\
            & KB & KB & mm$^2$ & \% CPU & \% DRAM & (\SI{}{\pico\joule}) & (\SI{}{\milli\watt}) & KB & KB & mm$^2$ & \% CPU & \% DRAM & (\SI{}{\pico\joule}) & (\SI{}{\milli\watt}) \\\hline
        \textbf{\X{}} &\textbf{10.63}& \textbf{8.30} & \textbf{0.04} & \textbf{0.02} &\textbf{-}&\ext{\textbf{25.98}} & \ext{\textbf{12.22}} & \textbf{85.00} & \textbf{66.41} & \textbf{0.25} & \textbf{0.11} & \textbf{-} &\textbf{36.87} & \textbf{50.\ext{54}} \\
        \quad Row ID Table & - & 5.64 & 0.01 & <0.01 & - & 1\ext{2.85} & 6.\ext{61} & - & 45.16 & 0.12 & 0.05 & - & 20.64 & 27.\ext{56}\\
        \quad Row Activation Counter Table & - & 2.66 & 0.02 & <0.01 & - & 11.13 & 4.66 & - & 21.25 & 0.06 & 0.03 & - & 11.66 & 15.53\\
        \quad Sibling Activation Vector & 10.63 & - & 0.01 & {<0.01} & - & \ext{2.00} & 0.95 & 85.00 & - & 0.07 & 0.03 & - & 4.57 & 7.44\\
        \textbf{PARA~\cite{kim2014flipping} }   & \textbf{-} & \textbf{-} & \textbf{-} & {<0.01} & \textbf{-} & \textbf{-} & \textbf{-} & \textbf{-} & \textbf{-} &\textbf{-} &<0.01 & \textbf{-} &\textbf{-} & \textbf{-}  \\
        \textbf{Graphene~\cite{park2020graphene} } & \textbf{-} & \textbf{286.51} & \textbf{0.81} & \textbf{0.35} & \textbf{-} &\textbf{87\ext{3.38}} & \textbf{18\ext{7.98}} & \textbf{-} & \textbf{2037.09} & \textbf{5.68} & \textbf{2.43} & \textbf{-} &\textbf{1042.49} & \textbf{1385.52} \\
        \textbf{Hydra~\cite{qureshi2022hydra} } & \textbf{61.56} & \textbf{-} & \textbf{0.10} & \textbf{0.04} & \textbf{-} & \textbf{43.07} & \textbf{24.\ext{17}} & \textbf{56.5} & \textbf{-} & \textbf{0.07} & \textbf{0.03} & \textbf{-} & \textbf{40.2\ext{6}} & \textbf{23.\ext{21}}\\
        \textbf{REGA~\cite{marazzi2023rega} } & \textbf{-} & \textbf{-} & \textbf{-} & \textbf{-} & \textbf{2.06} &\textbf{-} & \textbf{-} & \textbf{-} & \textbf{-} & \textbf{-} & \textbf{-} & \textbf{2.06} &\textbf{-} & \textbf{-} \\
        \hline
    \end{tabular}
    }
    \label{tab:area_cost_analysis}
\end{table*}
\end{savenotes}

\X{} is implemented in the memory controller{. It} does \emph{not} require \emph{any} modifications to existing DRAM chips. 

\noindent
\yct{\textbf{\iey{Key} Components. }}\yct{\X{}'s hardware implementation consists of two components: i)~the \X{} counter table, and ii)~the spillover counter.}
\yct{The \X{} counter table contains: i)~the Row ID Table (RIT), ii)~the RAC Table (RACT), and iii)~the SAV Table (SAVT).}
\iey{To efficiently track the number of activations, \X{} has to \yct{frequent}ly search and update the RIT and the RACT. Thus, we implement RIT and RACT using content-addressable memory (CAM) arrays. We implement SAVT as an SRAM array {since} \X{} does \emph{not} search SAVT entries. A register stores the spillover counter's value.}

\noindent
\yct{\textbf{Performing Preventive Refresh. }}\yct{Since the {standard} refresh (\gls{ref}) command is row-address-agnostic in DRAM standards~\cite{jedec2017ddr4,jedec2020ddr5,liu2012raidr}, \X{} cannot use standard refresh commands to refresh detected victim rows. To remain compatible with existing DRAM chips and interface standards, \X{} performs a preventive refresh operation by accessing {a} victim row once using \gls{act} and $PRE$ commands. When a tracked row's RAC value reaches PRT, \X{} performs preventive refresh operations to victim rows in all banks. 
\X{} prioritizes preventive refreshes over other memory requests: the memory controller does \emph{not} serve any memory request \emph{{to the same bank}} until the victim rows are preventively refreshed.}

\subsection{Area Overhead}
\label{sec:evaluation-area}

We evaluate \X{}' chip area{, static power, and memory array access energy} using CACTI~\cite{cacti}. Table~\ref{tab:area_cost_analysis} shows an area{, power, and energy} cost analysis of \X{} along with four other state-of-the-art RowHammer mitigation mechanisms~\cite{qureshi2022hydra,marazzi2023rega,park2020graphene,kim2014flipping} ({their configuration details are explained in \secref{sec:evaluation:methodology})}. We perform this analysis at \gls{nrh} values of 1000 and 125. Table~\ref{tab:params} shows the key parameters of \X{} for different \gls{nrh} values.

%
\begin{table}[!ht]
\caption{\X{} Parameters}
\resizebox{\linewidth}{!}{
\begin{tabular}{clccccc}
\toprule
\textbf{Term}   & \textbf{Definition}          &  & \multicolumn{4}{c}{\textbf{Value}} \\ \cmidrule(lr){1-1}\cmidrule(lr){2-2}\cmidrule(lr){4-7}
\textbf{\gls{nrh}}   & RowHammer threshold          &           & 1000    & 500    & 250   & 125   \\
\textbf{$PRT$}       & Preventive refresh threshold &           & 500    & 250    & 125   & 62   \\
\textbf{{$RCT$}}       & {Refresh cycle threshold} &           & 498    & 248    & 123   & 60   \\
\textbf{$N_{entries}$} & Number of table entries      &           & 2720    & 5440     & 10880   & 21760    \\
\textbf{$S_{SAV}$}   & Bit-length of a SAV entry    &           & 32      & 32     & 32    & 32    \\
\textbf{$S_{RID}$}   & Bit-length of a Row ID entry &           & 17      & 17     & 17    & 17    \\
\textbf{$S_{RAC}$}   & Bit-length of a RAC entry    &           & 10      & 9      & 8     & 7     \\ \hline
\end{tabular}
}
\label{tab:params}
\end{table}

All three \X{} hardware structures (Row ID Table, RACT, and SAVT) contain $N_{entries}$ entries. 
At a near future \gls{nrh} of 1000,
we estimate \X{}'s overall area overhead {to be} $0.04mm^2$ per DRAM channel for a dual-rank system. \X{} consumes approximately 0.02\% of the chip area of a high-end Intel Xeon processor with four memory channels~\cite{wikichipcascade}. At a low \gls{nrh} of 125, \X{}'s estimated chip area cost increases to $0.25mm^2$, taking up \emph{only} approximately 0.11\% of the same processor's area.

\noindent
\textbf{Area Comparison.} At an \gls{nrh} of 1000, \X{} takes up \param{20.25}$\times\!$ and \param{2.50}$\times\!$ smaller chip area than Graphene~\cite{park2020graphene} and Hydra~\cite{qureshi2022hydra}, respectively. Graphene's area overhead is larger than other mechanisms because it implements a large number of counters (e.g., 2720 per bank, 87040 in total for a dual-rank DDR4 system). REGA~\cite{marazzi2023rega} takes \param{2.06}\% DRAM chip area to implement. Compared to \X{}'s {memory controller chip area requirement}, REGA requires a larger {DRAM} chip area. PARA~\cite{kim2014flipping} does not maintain any state, thus it has \emph{no} significant area overhead.

\nb{We repeat our area overhead analysis for future DRAM chips by scaling the RowHammer threshold down to 125. Although \X{}'s area overhead increases as it implements a larger number of \X{} counters {at the lower \gls{nrh}}, {\X{} still} consumes a relatively small \param{0.11}\% processor chip area at an \gls{nrh} of 125. \X{} requires \param{22.72}$\times{}\!$ less chip area to implement than Graphene. \X{}'s area overhead at this very low \gls{nrh} is \param{3.57}$\times{}\!$ that of Hydra's\atbcr{, however, Hydra incurs up to a very large \param{85.42\%} performance overhead for 8-core memory-intensive workloads at the same \gls{nrh} (see~\secref{sec:evaluation-performance-energy})}. Hydra's chip area overhead reduces with decreasing \gls{nrh} as it requires counters with fewer bits of storage each.
We conclude that \X{}'s chip area requirement scales better than Graphene's and that \X{}'s area requirement at low \gls{nrh} is closer to the most area-efficient state-of-the-art mitigation mechanism, Hydra.}

\noindent
{\textbf{Energy and Static Power Comparison.} \nb{For an \gls{nrh} of 1000, \X{} has \param{3\ext{3.62}}$\times{}\!$ and \param{1.\ext{66}}$\times{}\!$ smaller access energy than Graphene and Hydra, respectively. {At the same \gls{nrh},} \X{} consumes $12.\ext{22}mW$ of static power, which is \param{15.\ext{38}}$\times{}\!$ and \param{1.9\ext{8}}$\times{}\!$ smaller than Graphene and Hydra's static power consumptions. As \gls{nrh} reduces to 125, \X{}'s static power and access energy scale more efficiently (similarly to Hydra) compared to Graphene, where Graphene has \param{2\ext{8.27}}$\times{}\!$ and \param{27.\ext{41}}$\times{}\!$ the access energy and static power of \X{}, respectively.}}

\subsection{Latency Analysis}
We implement \X{} in Verilog HDL and use Synopsys DC~\cite{synopsys} to evaluate \X{}'s latency impact on memory accesses. According to our Verilog model, \X{} \revcommon{needs} \SI{1.22}{\nano\second} \revcommon{to update the \X{} counter of an activated DRAM row}. This latency \revcommon{overlaps with the latency of regular memory controller operations as} it is smaller than \gls{trrd} (e.g., \SI{2.5}{\nano\second} in DDR4~\cite{micron2014ddr4,jedec2017ddr4}).


\section{Evaluation Methodology}
\label{sec:evaluation:methodology}

\nb{We evaluate \mech{}'s performance and energy consumption with Ramulator \cite{kim2016ramulator, ramulatorgithub}, a cycle-accurate DRAM simulator, and DRAMPower\cite{drampower}. We specify our simulated system's configuration in Table \ref{configs}.}

\begin{table}[h]
\centering
\caption{Simulated System Configuration}
\label{configs}
\resizebox{\linewidth}{!}{
\begin{tabular}{ll}
\hline
\textbf{Processor}                                                   & \begin{tabular}[c]{@{}l@{}} 1 or 8 cores, 3.\revcommon{6}GHz clock frequency,\\ 4-wide issue, 128-entry instruction window\end{tabular}  \\ \hline
\textbf{DRAM}                                                        & \begin{tabular}[c]{@{}l@{}}DDR4, 1 channel, 2 rank/channel, 4 bank groups,\\ 4 banks/bank group, 128K rows/bank, 3200 MT/s\end{tabular}  \\ \hline
\begin{tabular}[c]{@{}l@{}}\textbf{Memory Ctrl.}\end{tabular} & \begin{tabular}[c]{@{}l@{}}64-entry read and write requests queues,\\Scheduling policy: FR-FCFS~\cite{rixner2000memory,zuravleff1997controller} \\with a column cap of 16~\cite{mutlu2007stall},\\Address mapping: MOP\cite{kaseridis2011minimalistic,zhang2000permutation}\\\SI{45}{\nano\second} $tRC$, \SI{7.9}{\micro\second} $tREFI$, \SI{64}{\milli\second} $tREFW$\\ \SI{64}{\milli\second} \X{} reset period\end{tabular}   \\ \hline
\textbf{Last-Level Cache}& \begin{tabular}[c]{@{}l@{}} 2 MiB per core \end{tabular}  \\ \hline
\end{tabular}}
\end{table}

\begin{figure}[h]
    \centering
    \includegraphics[width=0.8\linewidth]{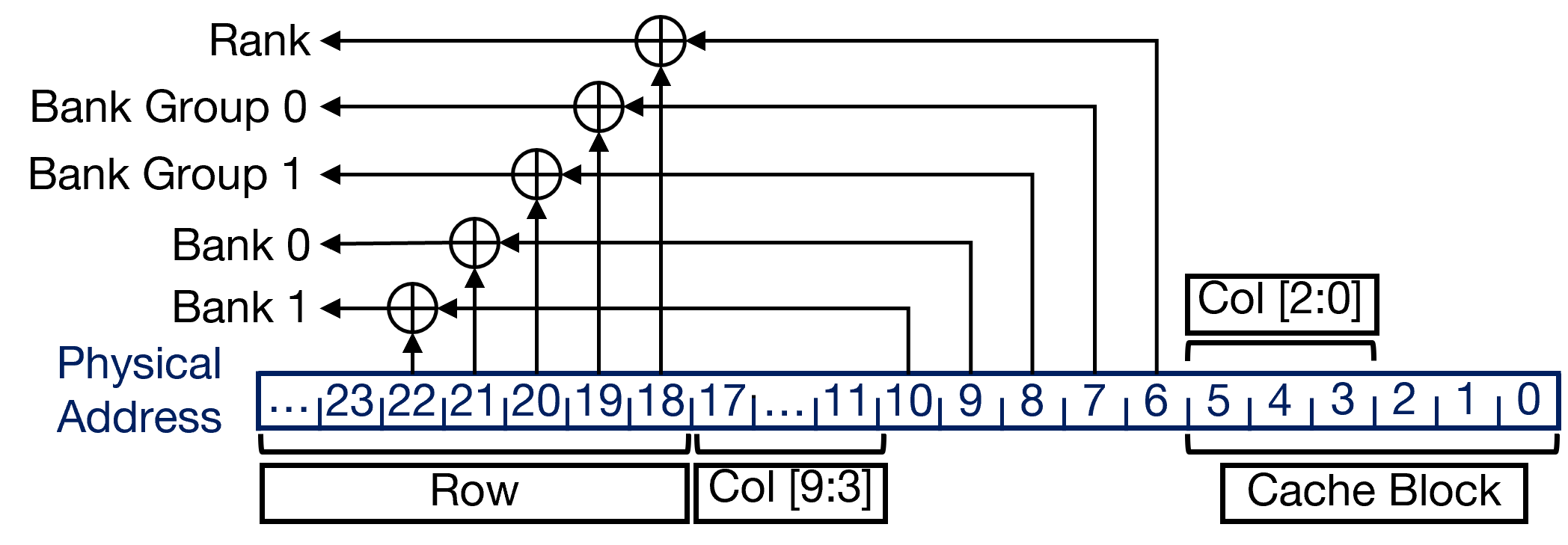}
    \caption{Simulated address mapping}
    \label{fig:address-mapping}
\end{figure}

\noindent
\textbf{Address Mapping.} \figref{fig:address-mapping} depicts our address mapping scheme. We use an address mapping scheme that interleaves consecutive cache blocks in the physical address space between different DRAM banks. \ext{Appendix~\ref{appendix:mapping} evaluates \X{} using similar mapping schemes that interleave sets of 2, 4, and 8 cache blocks in the physical address space between different DRAM banks.}

\noindent
\textbf{Comparison Points.} We compare \X{} to a baseline system with no RowHammer mitigation and to \param{four} state-of-the-art RowHammer mitigation mechanisms: (1) Graphene~\cite{park2020graphene} implements per bank counters to track the possible aggressor rows using Misra-Gries algorithm~\cite{misra1982finding}. When a counter value exceeds a threshold value, Graphene issues preventive refreshes to the victim rows. (2) Hydra~\cite{qureshi2022hydra} implements a group count table to track activations for a group of rows and a row count table to track per row activations. \revcommon{We configure Hydra such that all rows in a row group have their row count table entries reside in two consecutive cache blocks (64 bytes each), as described in~\cite{qureshi2022hydra}.} The row count table is stored in the DRAM and cached in the memory controller. Hydra performs preventive refresh operations when a counter exceeds a threshold value. (3) REGA~\cite{marazzi2023rega} augments DRAM design such that one or more victim rows can be refreshed when a DRAM row is activated. REGA tunes its protection guarantees by changing the default \gls{trc} value. A smaller \gls{trc} allows REGA to refresh more rows concurrently with a DRAM row activation, at the cost of increased access latency. To simulate REGA, we modify \gls{trc} as described in~\cite{marazzi2023rega}. (4) PARA~\cite{kim2014flipping} protects against RowHammer by performing probabilistic adjacent row activation. When a row is closed {(i.e., when the memory controller issues a \gls{pre})}, PARA issues preventive refreshes to the adjacent rows based on a probability threshold. We tune the probability threshold of PARA for a target failure probability of $10^{-15}$ within a \SI{64}{\milli\second} as in prior work~\cite{yaglikci2021blockhammer}. {Table~\ref{table:other-mechanisms} summarizes the key configuration parameters of the evaluated state-of-the-art mechanisms.}

\begin{table}[!ht]
\caption{\atbcr{Key configuration parameters of state-of-the-art mechanisms}}
\label{table:other-mechanisms}
\resizebox{\linewidth}{!}{
\begin{tabular}{l|l||c|c|c|c}
\toprule
\textbf{Mechanism}                 & \textbf{Configuration Parameter}                             & \multicolumn{4}{c}{\textbf{Value}}                    \\
\toprule
\toprule
\textbf{All mechanisms}             & RowHammer Threshold                         & 1000      & 500       & 250       & 125      \\
                          \midrule
\multirow{3}{*}{Graphene} & Number of table entries          & 2720      & 5440      & 10880     & 21760    \\
\cline{2-6}
                          & Threshold for aggressor tracking & 500       & 250       & 125       & 63       \\
                          \cline{2-6}
                          & Reset window                     & \multicolumn{4}{c}{\SI{64}{\milli\second}}                    \\
\midrule
\midrule
\multirow{6}{*}{Hydra}    & Row group size                   & \multicolumn{4}{c}{128 rows} \\
\cline{2-6}
                          & Row count table entry size       & 2B        & \multicolumn{3}{c}{1B}           \\
                          \cline{2-6}
                          & Row count cache size             & \multicolumn{4}{c}{4K entires per DRAM rank} \\
                          \cline{2-6}
                          & Group count table threshold      & 400       & 200       & 100       & 50       \\
                          \cline{2-6}
                          & Tracking threshold               & 500       & 250       & 125       & 63       \\
                          \cline{2-6}
                          & Periodic reset                   & \multicolumn{4}{c}{\SI{64}{\milli\second}}                    \\
\midrule
\midrule
REGA                      & Row cycle time ($t_{RC}$)        & \SI{45.0}{\nano\second}     & \SI{62.5}{\nano\second}   & \SI{97.5}{\nano\second}   & \SI{167.5}{\nano\second} \\
\midrule
\midrule
PARA                      & Probability threshold            & 0.034     & 0.067     & 0.129     & 0.241   \\
\toprule
\end{tabular}
}
\end{table}

\noindent
\ste{\textbf{Workloads.} We evaluate \reviii{\param{62}} single-core and \reviii{\param{62}} {homogeneous multi-programmed} 8-core workload\omcri{s} from \nb{five} benchmark suites: \agy{SPEC CPU2006~\cite{spec2006}}, SPEC CPU2017~\cite{spec2017}, \nb{TPC~\cite{tpcweb}, MediaBench~\cite{fritts2009media}, and YCSB~\cite{ycsb}}.
Based on the row buffer misses-per-kilo-instruction (RBMPKI), we group the applications into three categories, which Table \ref{table:workloads} describes: (1) L (low memory-intensity, RBMPKI $\in [0,2)$), (2) M (medium memory-intensity, RBMPKI $\in [2,10)$), (3) H (high memory-intensity, RBMPKI $\in [10+)$). To do so, we obtain
the RBMPKI values of the applications by analyzing each application's SimPoint \cite{calder2005simpoint} traces (200M instructions). {All of these traces are open-sourced~\cite{self.github}.}}

\ste{
\begin{table}[ht]
\Copy{C3table}{
\caption{Evaluated {single-core} workloads}
\label{table:workloads}
\footnotesize
\centering
\resizebox{\linewidth}{!}{
\begin{tabular}{|l||l|}
\hline
\multicolumn{1}{|c||}{\textbf{RBM\revcommon{P}KI}}                      & \multicolumn{1}{c|}{\textbf{Workloads}}                                                                                                                                                                                                                                                                             \\ \hline \hline
\begin{tabular}[c]{@{}l@{}}$[10+)$ \\ (High)\end{tabular} & \begin{tabular}[c]{@{}l@{}}
519.lbm, 459.GemsFDTD, 450.soplex, h264\_decode, \\
520.omnetpp, 433.milc, 434.zeusmp, bfs\_dblp, \\
429.mcf, 549.fotonik3d, 470.lbm, bfs\_ny,\\
bfs\_cm2003, 437.leslie3d, \reviii{gups}
\end{tabular}      
\\ \hline
\begin{tabular}[c]{@{}l@{}}$[2, 10)$\\ (Medium)\end{tabular}  & \begin{tabular}[c]{@{}l@{}}
510.parest, 462.libquantum, tpch2, wc\_8443, \\
ycsb\_aserver, 473.astar, jp2\_decode, 436.cactusADM,\\ 
557.xz, ycsb\_cserver, ycsb\_eserver, 471.omnetpp, \\
483.xalancbmk, 505.mcf, wc\_map0, jp2\_encode, \\
tpch17, ycsb\_bserver, tpcc64, 482.sphinx3
\end{tabular}                                                                                                                      \\ \hline
\begin{tabular}[c]{@{}l@{}}$[0, 2)$\\ (Low)\end{tabular}     & \begin{tabular}[c]{@{}l@{}}
502.gcc, 544.nab, h264\_encode, 507.cactuBSSN,\\
525.x264, ycsb\_dserver, 531.deepsjeng, 526.blender, \\
435.gromacs, 523.xalancbmk, 447.dealII, 508.namd, \\
538.imagick, 445.gobmk, 444.namd, 464.h264ref, \\
ycsb\_abgsave, 458.sjeng, 541.leela, tpch6, \\
511.povray, 456.hmmer, 481.wrf, grep\_map0, \\
500.perlbench, 403.gcc, 401.bzip2
\end{tabular} \\ \hline
\end{tabular}
}
}
\end{table}
}

\section{Evaluation}
\label{sec:evaluation}

{We 1) analyze \X{}'s system performance and DRAM energy overheads and compare \X{}'s system performance and DRAM energy overheads to state-of-the-art mitigation mechanisms (\secref{sec:evaluation-performance-energy}), 2) show the effect of the number of banks in the system on \X{}'s performance, and 3) analyze \X{}'s performance under RowHammer attacks.}

\subsection{System Performance and DRAM Energy}
\label{sec:evaluation-performance-energy}
\begin{figure}[h]
    \centering
    \includegraphics[width=\linewidth]{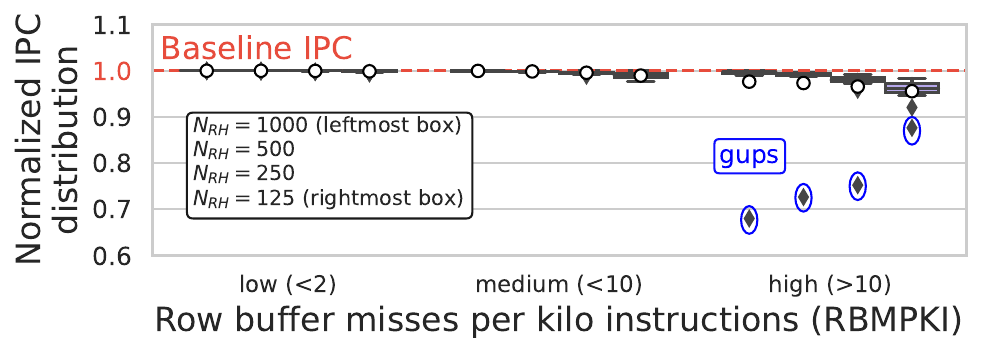}
    \caption{Performance of single-core applications for four different RowHammer thresholds (higher is better).}
    \label{fig:perf_single_core_small}
\end{figure}

\noindent
\textbf{System Performance Overhead.}
\figref{fig:perf_single_core_small} presents the performance (in instructions per cycle) of all single-core workloads (grouped into three categories and sorted based on RBMPKI; see Table~\ref{table:workloads}) for four different near-future and {very} low RowHammer thresholds when executed on a system that uses \X{}, normalized to a baseline system that does not employ any RowHammer mitigation mechanism.\footnote{\revcommon{Appendix~\ref{appendix:single-core-extended} plots the performance of \X{} for each workload.}}

We make \reviii{{two}} major observations from~\figref{fig:perf_single_core_small}. First, \X{} induces minor system performance overhead for all evaluated single-core workloads at a near-future \gls{nrh} of 1000. At such \gls{nrh}, \X{} incurs \emph{only} \reviii{\param{0.58}}\% (\reviii{\param{32.00}}\%) slowdown on average (at maximum) across all workloads. \X{} increases the average memory latency experienced by application memory requests by \reviii{\param{1.87}}\% on average across all workloads {(not shown)} due to preventive refreshes. Second, \X{} can efficiently prevent RowHammer bitflips even at {very} low \gls{nrh}. At {such} a future \gls{nrh} of 125, \X{} induces \reviii{\param{1.45}}\% (\param{12.37}\%) slowdown on average (at maximum) across all workloads. At {this} \gls{nrh}, the average memory latency increases by \param{2.72}\% across all workloads on average due to preventive refresh operations. {We attribute the increasing trend in \X{}'s average slowdown to the larger number of preventive refresh operations performed by \X{} at lower \gls{nrh} values. More workloads hammer more rows more times in a refresh window as \gls{nrh} reduces, which leads to both i) \X{} row activation counters (RACs) incrementing faster and ii) more \X{} RACs reaching the preventive refresh threshold earlier (and \X{} performing costly preventive refresh operations).}
{In contrast to the trend in the average slowdown, we observe that \X{} induces a smaller maximum slowdown as \gls{nrh} reduces.}
{This is because \X{} implements more row activation counters at lower \gls{nrh} values and a very memory-intensive random access workload (e.g., gups) \emph{cannot} quickly exhaust all \X{} RACs and increases the spillover counter value to the refresh cycle threshold slower (than at higher \gls{nrh} values) such that \X{} less frequently performs refresh cycles (\secref{sec:mechanism-operation}).}

\begin{figure}[h]
    \centering
    \includegraphics[width=\linewidth]{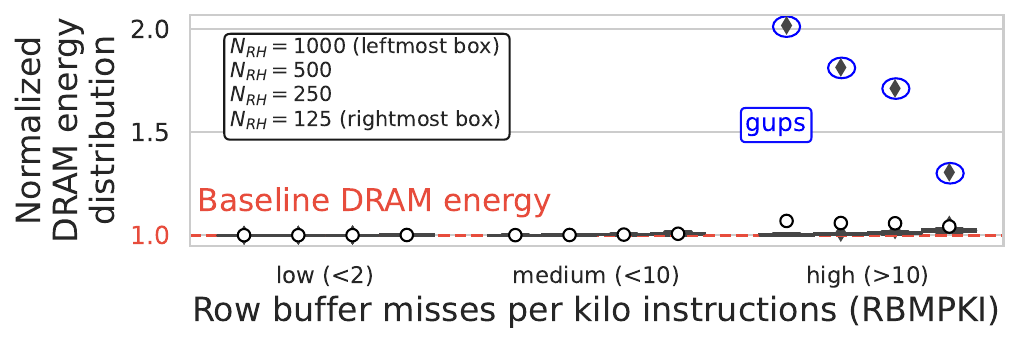}
    \caption{DRAM energy for single-core applications for four different RowHammer thresholds (lower is better).}
    \label{fig:energy_single_core_small}
\end{figure}

\noindent
\textbf{DRAM Energy Overhead.}
\figref{fig:energy_single_core_small} presents the DRAM energy consumption for all single-core workloads for four different RowHammer thresholds when executed on a system that uses \X{}, normalized to a baseline system that does not employ any RowHammer mitigation mechanism.\footnote{\revcommon{Appendix~\ref{appendix:single-core-extended} plots the normalized DRAM energy consumption of \X{} for each workload.}}

We make two key observations from~\figref{fig:energy_single_core_small}. First, \X{} induces minor DRAM energy overhead at \gls{nrh} {= 1000}. \X{} increases DRAM energy consumption by \emph{only} \reviii{\param{1.65}}\% \reviii{(2.02$\times{}$)} on average (at maximum) across all evaluated workloads. Second, \X{} increases DRAM energy consumption by \reviii{\param{1.27}}\% (\reviii{\param{30.46}}\%) on average (maximum) across all workloads at \gls{nrh}{=125}. We attribute the DRAM energy overheads to {i) increased DRAM activation, precharge, and command bus energy induced by the preventive refresh operations, and ii) increased DRAM background (standby) energy consumption due to increased execution time for applications.}

\begin{figure}[!ht]
    \centering
    \includegraphics[width=\linewidth]{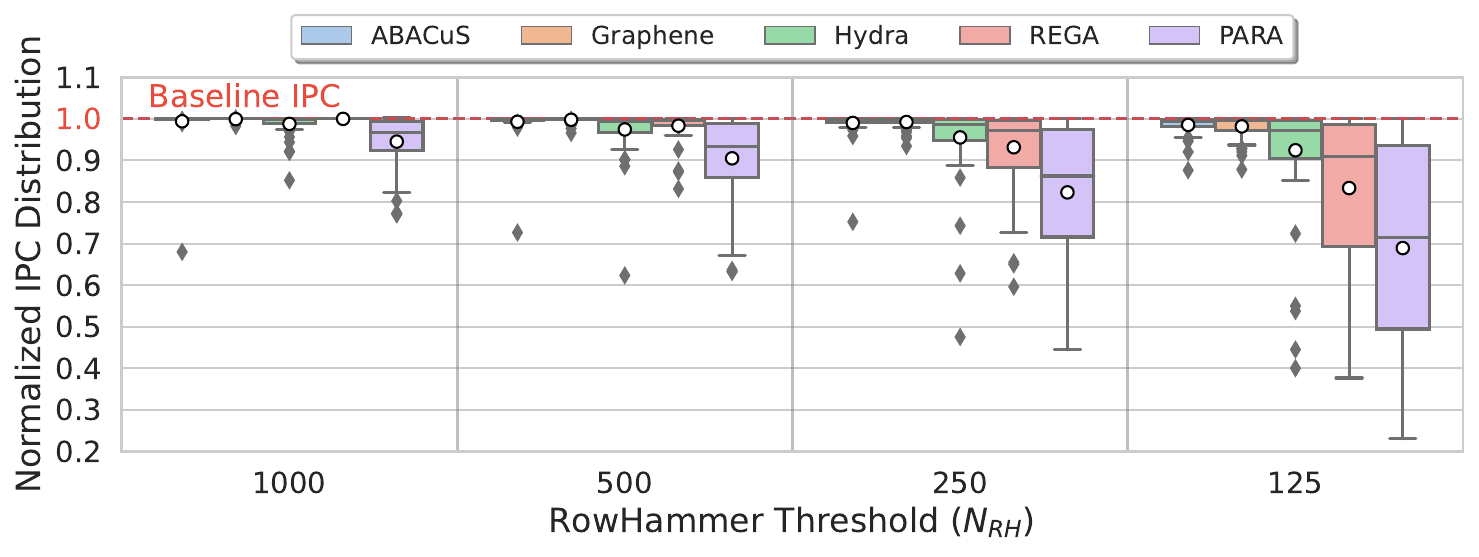}
    \caption{Performance comparison {of \X{} vs. state-of-the-art mitigation techniques} for single-core workloads at four different RowHammer thresholds.}
    \label{fig:perf_comparison_single_core}
\end{figure}

\noindent
\textbf{Performance Comparison.} \figref{fig:perf_comparison_single_core} presents the performance impact of \X{} and \param{four} state-of-the-art mechanisms on a single-core system for four different RowHammer thresholds, normalized to the baseline system.

We make \param{six} key observations based on~\figref{fig:perf_comparison_single_core}. \param{First}, \X{} outperforms Hydra, REGA, and PARA at RowHammer thresholds below 1000 and performs similarly to Graphene at all tested RowHammer thresholds on average across all workloads. \param{Second}, \X{} outperforms Hydra and PARA at \gls{nrh} {= 1000}. \param{Third}, REGA~\cite{marazzi2023rega} at \gls{nrh} {= 1000} does \emph{not} incur any performance overhead. At \gls{nrh} {= 1000}, REGA can hide the latency of a preventive refresh behind the latency of a DRAM row access (i.e., preventive refresh happens concurrently with a DRAM row access at the nominal \gls{trc} defined in the DDR4 standard~\cite{jedec2017ddr4}). However, REGA incurs increasingly higher overheads as \gls{nrh} reduces because REGA needs to perform multiple preventive refreshes on each DRAM row access. To perform 8 preventive refreshes on each DRAM row access at an \gls{nrh} of 125, REGA increases \gls{trc} from the nominal value of \SI{45.0}{\nano\second}~\cite{jedec2017ddr4} to \SI{167.5}{\nano\second},\footnote{With such a \gls{trc} value, REGA securely prevents RowHammer bitflips at an \gls{nrh} of 130. REGA \emph{cannot} be configured for an \gls{nrh} of 125 because refreshing 9 rows on each access would allow it to prevent RowHammer bitflips at an \gls{nrh} of 116. We evaluate REGA with a \gls{trc} of \SI{167.5}{\nano\second}, where we compare it against other mechanisms at an \gls{nrh} of 125 because an \gls{nrh} of 130 is closer to 125 than an \gls{nrh} of 116.} where REGA induces \reviii{\param{16.65}}\% performance overhead on average across all workloads as the average memory access latency increases.

\yct{Fourth}, PARA~\cite{kim2014flipping} performs the worst among all evaluated mechanisms. 
{PARA incurs \reviii{\param{5.47}}\% and \reviii{\param{31.08}}\% performance overhead\omcri{s} on average across all workloads at \gls{nrh} = 1000 and 125, respectively, because it performs many unnecessary refresh operations~\cite{yaglikci2021blockhammer,kim2020revisiting}.}

\yct{Fifth}, Hydra~\cite{qureshi2022hydra} incurs \reviii{\param{1.80}}\%, \param{3.33}\%, \param{5.70}\%, and \reviii{\param{9.75}}\% higher performance overheads than \X{} for \gls{nrh} of 1000, 500, 250, and 125, respectively, on average across all workloads. In addition to performing preventive refresh operations, Hydra also performs i) an \gls{act} and a \gls{wr} command when a counter in its row count cache (RCC) needs to be evicted to the row count table (RCT) in DRAM, and ii) an \gls{act} and a \gls{rd} command when a counter needs to be retrieved from the RCT and placed in the RCC. These operations incur additional performance overheads due to i) row buffer misses that interfere with application memory requests, and ii) DRAM banks being unavailable during RCC eviction and RCT access operations, on top of the overheads caused by preventive refresh operations. For example, at an \gls{nrh} of 125, {the Hydra-based system} {has} i) a row buffer miss rate \reviii{\param{6.22}}\% {larger than that of \X{}} and ii) {an} average memory latency experienced by application memory requests \reviii{\param{20.94}}\% higher than {that of} \X{} on average across all workloads.

\yct{Sixth}, Graphene~\cite{park2020graphene} incurs slightly higher performance overhead than \X{} on average across all workloads at an \gls{nrh} of \param{125}. Even though \X{}, compared to Graphene, performs \param{2.06}$\times\!$ more preventive refresh operations as \X{}'s shared activation counters reach the preventive refresh threshold faster, the amount of time where at least one DRAM bank is unavailable (for serving application memory requests) because of preventive refresh is an estimated \param{7.73}$\times\!$ higher in Graphene compared to \X{}. {Once an \X{} activation counter reaches \gls{nrh}, \X{} performs 64 preventive refresh operations (to all 64 victim rows in 32 banks of the rank) in \emph{quick succession}. The memory controller takes approximately \SI{170}{\nano\second}\footnote{{Calculated as the time it takes to activate and precharge two rows in the same bank ($2*tRC$) plus the number of banks multiplied by the minimum time between two successive \gls{act} commands to different banks in the same rank ($32*tRRD$).}} to issue all activate and precharge commands that make up a preventive refresh operation, leveraging bank-level parallelism. In contrast, issuing two preventive refresh operations to a single bank takes approximately \SI{90}{\nano\second}, an already large fraction of the time it takes to issue 64 preventive refresh operations. Keeping at least one DRAM bank unavailable for a longer total time} increases the critical path for application memory requests in Graphene by a larger amount than in \X{}. As such, the amount of time in which the processor \emph{cannot} execute instructions due to the re-order buffer being full is higher (i.e., pipeline stall cycles are higher) by \param{1.87}\% in Graphene compared to \X{} on average across all workloads.

\begin{figure}[!ht]
    \centering
    \includegraphics[width=\linewidth]{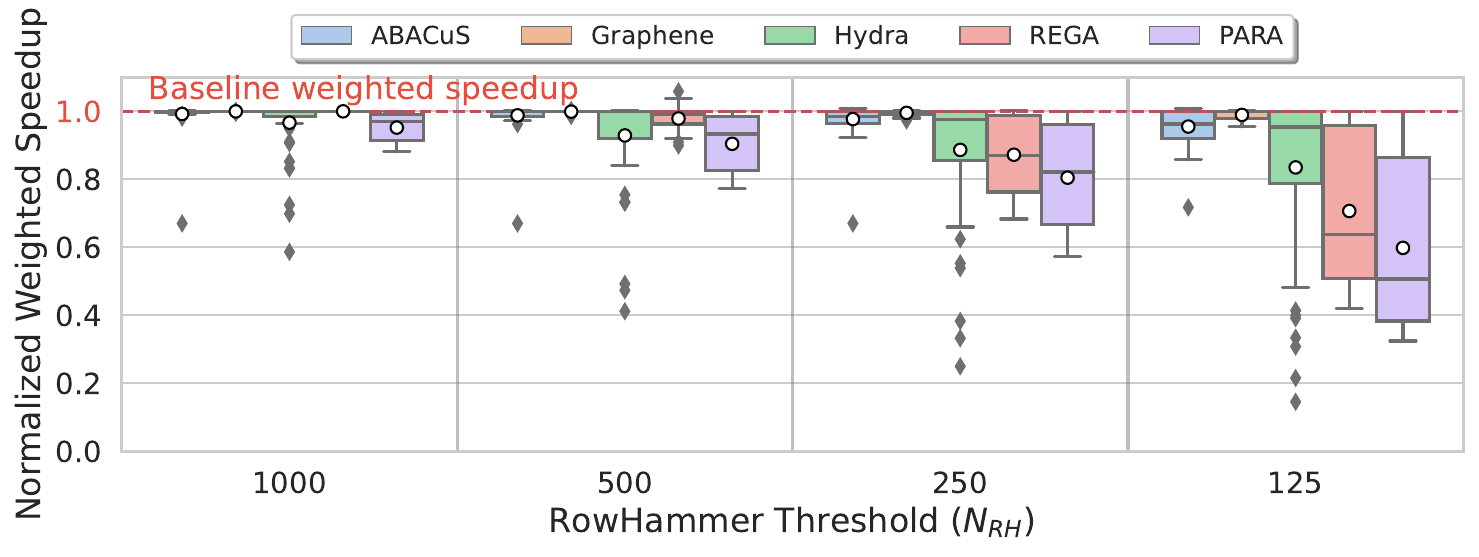}
    \caption{{Performance comparison for multi-programmed (8 core) workloads at four different RowHammer thresholds}}
    \label{fig:perf_comparison_multi_core}
\end{figure}

{\figref{fig:perf_comparison_multi_core} shows \X{} and \param{four} state-of-the-art mechanisms' performance impact in terms of weighted speedup~\cite{snavely2000symbiotic, eyerman2008systemlevel, michaud2012demystifying} for four different \gls{nrh} values on an eight-core system, normalized to the baseline system.\footnote{We simulate each eight-core workload until {all} cores execute 25M instructions to maintain a reasonable experiment time for eight-core workload simulations.} \ext{Appendix~\ref{appendix:heterogeneous-workloads} presents a figure that shows the weighted speedup for randomly generated heterogeneous 8-core workloads.}}

{We make \param{three} key observations from~\figref{fig:perf_comparison_multi_core}. 
\param{First}, \X{} induces small system performance overhead across all evaluated workloads and RowHammer thresholds. \X{} has \param{0.77}\%, \param{1.19}\%, \param{2.29}\%, and \param{4.48}\% performance overhead on average across all workloads for \gls{nrh} of 1000, 500, 250, and 125, respectively. \param{Second}, Hydra incurs \param{2.56}\% higher performance overhead than \X{} at an \gls{nrh} of 1K.
\param{Third}, \X{} outperforms Hydra, REGA, and PARA at an extreme RowHammer threshold of 125. At such \gls{nrh}, \X{} incurs \emph{only} \param{4.48}\% performance overhead, whereas Hydra, REGA, and PARA incur \param{16.49}\%, \param{29.31}\%, and \param{40.16}\% performance overhead on average across all workloads.}

\begin{figure}[!ht]
    \centering
    \includegraphics[width=\linewidth]{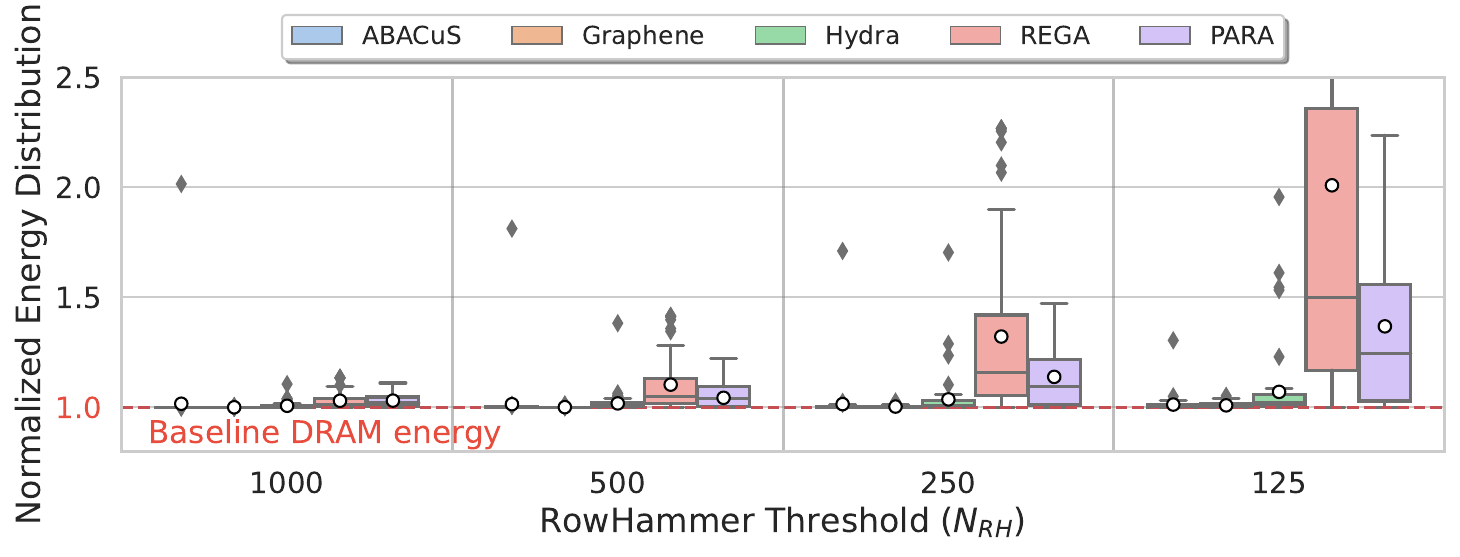}
    \caption{DRAM energy comparison for single-core workloads at four different RowHammer thresholds}
    \label{fig:energy_comparison_single_core}
\end{figure}

\noindent
\textbf{Energy Comparison.} \figref{fig:energy_comparison_single_core} presents the DRAM energy consumption of \X{} and \param{four} state-of-the-art mechanisms on a single-core system for four different RowHammer thresholds, normalized to the baseline system.

From~\figref{fig:energy_comparison_single_core}, we {make two observations. First,} \X{} induces smaller DRAM energy overhead than other evaluated mitigation mechanisms {(except Graphene)} on average across all workloads {for \gls{nrh}<1000. {Second,} at an \gls{nrh} of 1000, \X{} induces \param{1.34}\% and \param{1.36}\% smaller DRAM energy overheads than REGA and PARA, respectively, because} 
{REGA preventively refreshes one row with {every} DRAM row activation (at \gls{nrh} = 1000) and PARA performs many unnecessary refresh operations. \X{} induces \param{1.66\%} average (\param{2.02}$\times\!$ maximum) DRAM energy overhead at this \gls{nrh}, which is close to Hydra's 0.73\% average (\param{1.11}$\times\!$ maximum) DRAM energy overhead.}

\begin{figure}[!ht]
    \centering
    \includegraphics[width=\linewidth]{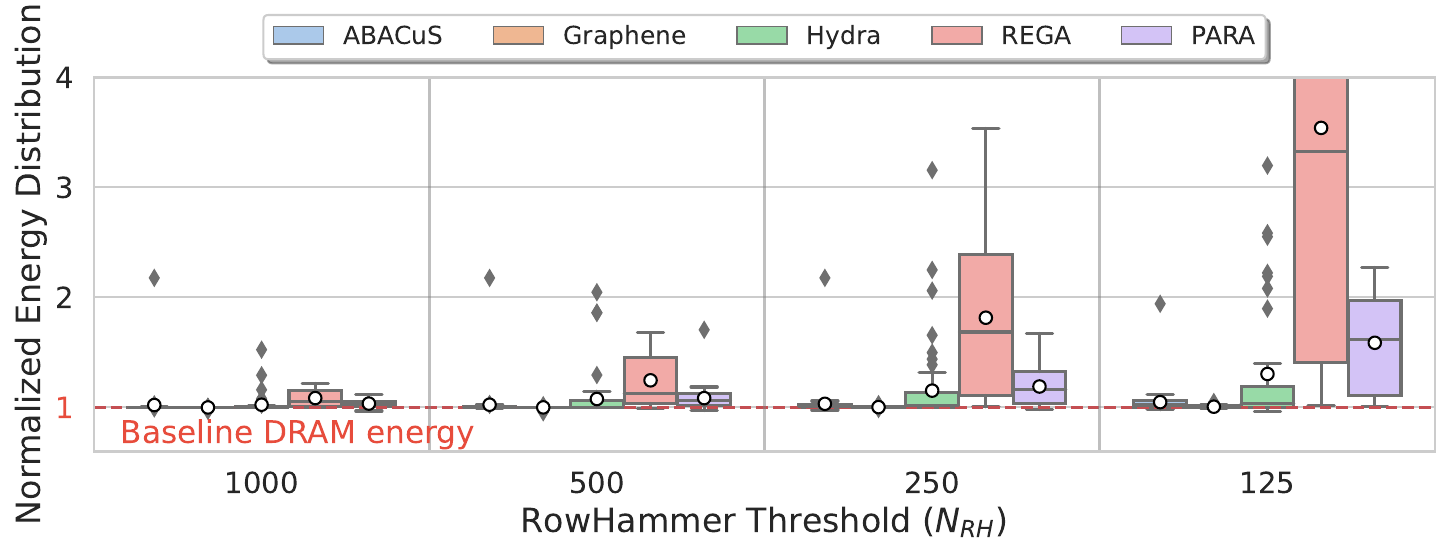}
    \caption{{DRAM energy comparison for multi-programmed (8 core) workloads at four different RowHammer thresholds}}
    \label{fig:energy_comparison_multi_core}
\end{figure}

{\figref{fig:energy_comparison_multi_core} shows the DRAM energy consumption of \X{} and \param{four} state-of-the-art mechanisms for four different \gls{nrh} on an eight-core system, normalized to the baseline system.}

{From~\figref{fig:energy_comparison_multi_core}, we observe that \X{} induces \param{2.12\%}, \param{2.44\%}, \param{3.25\%}, and \param{4.76\%} DRAM energy overhead at \gls{nrh} = 1000, 500, 250, and 125, respectively. \atbcr{At a very low \gls{nrh} = 125,} \X{}'s DRAM energy overhead is \param{19.64\%}, \param{70.41\%}, and \param{33.99\%} smaller than Hydra, REGA, and PARA, on average across all evaluated workloads. Graphene induces \param{3.95\%} lower DRAM energy overhead than \X{} \atbcr{at \gls{nrh} = 125}.}

{\textbf{Summary.} We conclude that} {\X{} induces small system performance and DRAM energy overheads on average across all tested single-core and multi-core workloads for \gls{nrh} = 1000, 500, 250, and 125. \X{}'s performance and DRAM energy overheads are closer to the most-performance-efficient state-of-the-art mechanism (Graphene~\cite{park2020graphene}). \X{} outperforms and consumes less DRAM energy than the most-area-efficient state-of-the-art (counter-based) mechanism (Hydra~\cite{qureshi2022hydra}).}

\subsection{\ext{Sensitivity to Table Entries ($N_{entries}$)}}
\begin{figure}[!h]
    \centering
    \includegraphics[width=0.95\linewidth]{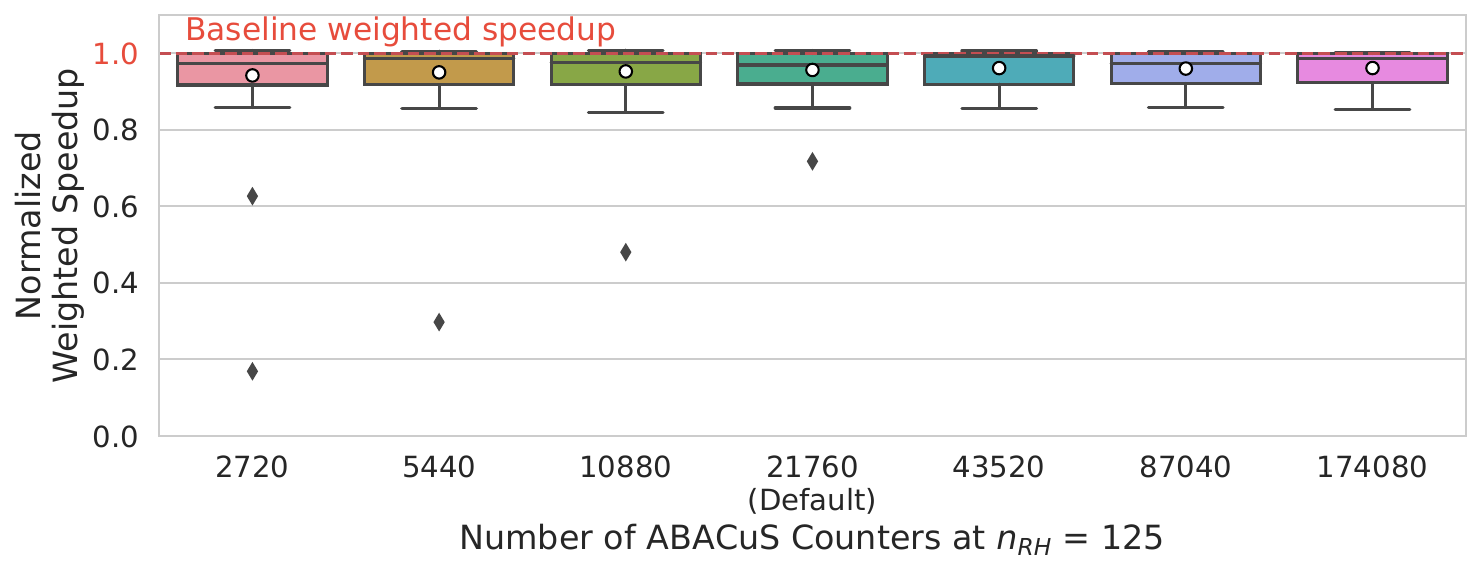}
    \caption{\ext{Normalized performance (distribution) of single-core applications with varying numbers of \X{} table entries (i.e., \X{} counters) at \gls{nrh} = 125.}}
    \label{fig:sensitivity-nentries}
\end{figure}

\ext{We demonstrate how the number of \X{} table entries ($N_{entries}$, or the number of \X{} counters in the counter table) affects system performance. \figref{fig:sensitivity-nentries} depicts the normalized weighted speedup (y-axis) of all 62 homogeneous 8-core workloads for different numbers of \X{} table entires (x-axis) in an \X{} system configured for \gls{nrh} = 125.} 

\ext{We make two major observations. First, implementing fewer than the default number (21760) of table entries marginally increases the average and significantly increases the maximum performance overheads. For example, at $N_{entries}$ = 21760, \X{} incurs 4.48\% average and 28.24\% maximum performance overheads. \X{}' average and maximum performance overheads increase to 5.79\% and 83.10\%, respectively, at $N_{entries}$ = 2720. Random access, memory-intensive workloads cause refresh cycles more frequently in \X{} as $N_{entries}$ becomes smaller as the spillover counter keeps track of more rows. Second, conversely, implementing more than the default number of table entries reduces \X{}' performance overheads. For example, \X{}'s average and maximum performance overheads at $N_{entries}$ = 174080 are 3.86\% and 14.64\%, respectively. However, increasing $N_{entries}$ beyond 43520 has diminishing benefits.}

\ext{We conclude that a larger number of \X{} table entries reduces \X{}' performance overheads.\footnote{\ext{More table counters come at the cost of more chip area overhead.}}}

\subsection{Sensitivity to Number of Banks}
We run 16-, 32-, and 64-bank (1-, 2-, and 4-rank) simulations using \X{} and the baseline system. We observe that \X{} can prevent RowHammer bitflips with low overhead in systems that use memory modules with different numbers of {banks (}ranks). At \gls{nrh} = 125, \X{} incurs \reviii{\param{1.58}}\%, \reviii{\param{1.50}}\%, and \reviii{\param{2.60}}\% performance overheads for 16-, 32-, and 64-bank configurations, respectively, on average (geometric mean) across all evaluated single-core workloads.

\begin{figure}[!h]
    \centering
    \includegraphics[width=0.95\linewidth]{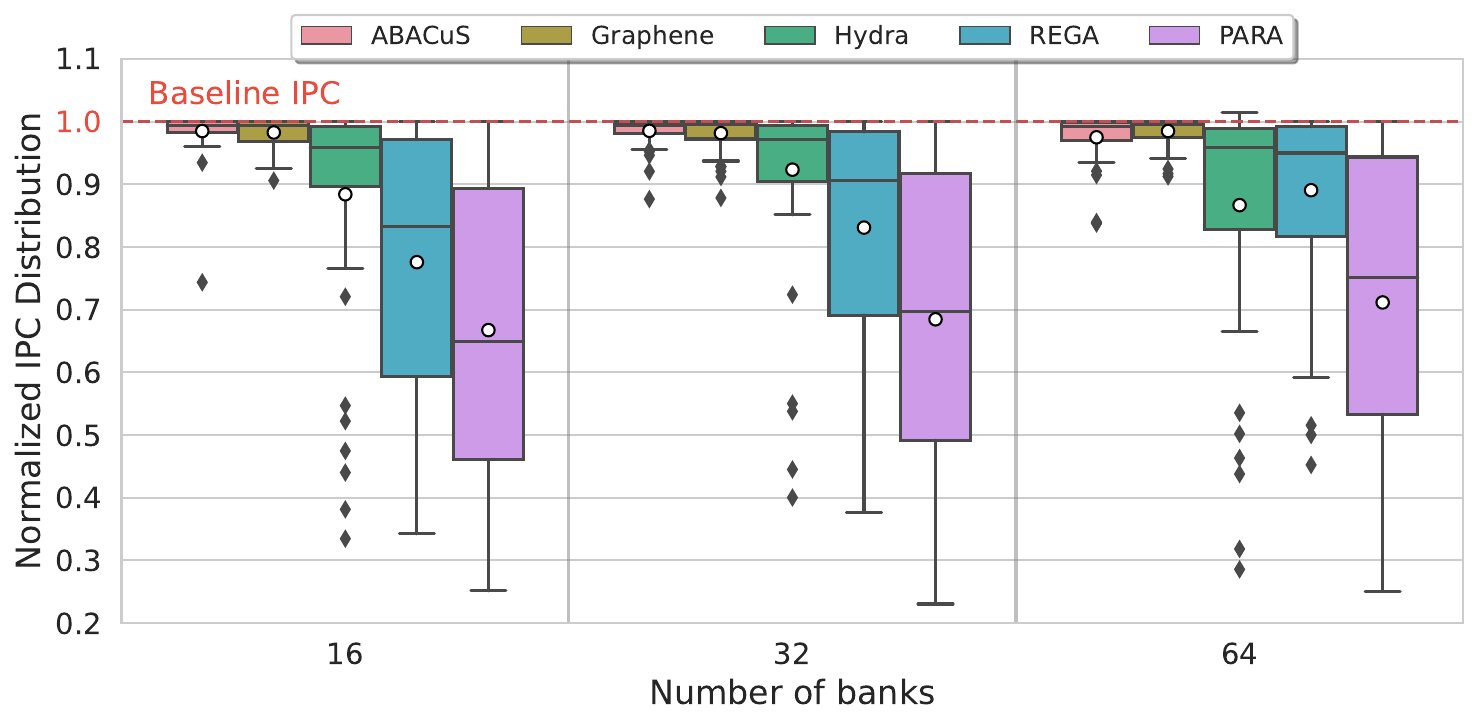}
    \caption{\ext{Normalized performance (distribution) of single-core applications with varying numbers of DRAM banks in the system for \X{} and four state-of-the-art mechanisms (\gls{nrh} = 125).}}
    \label{fig:sensitivity-banks}
\end{figure}

\ext{\figref{fig:sensitivity-banks} shows the performance (in instructions per cycle) distribution of all 62 single-core workloads with varying numbers of DRAM banks (16, 32, and 64) on the x-axis, for \X{} and four state-of-the-art mitigation mechanisms (different boxes) normalized to the baseline system equipped with the same number of banks.}

\ext{We make two major observations. First, \X{}'s average performance overhead does \emph{not} correlate strongly with the number of banks. We observe that the tail of the normalized performance distribution and the average normalized performance increase and decrease as the number of banks increases from 16 to 64. We attribute the performance improvement from 16 banks to 32 banks to workload access patterns changing (as the number of banks changes) in a way that reduces the preventive refresh operations performed by \X{}. For example, the 401.bzip2 workload triggers \param{6912} and \param{704} preventive refresh operations, resulting in \param{25.64}\% and \param{12.37}\% system performance overheads at 16 banks and 32 banks, respectively. We attribute the performance degradation when moving from 32 to 64 banks to \X{}' increased number of preventive refreshes. More sibling rows share the same \X{} counter, and this causes \X{} counter values to increase more aggressively (compared to when there are fewer banks). We observe that for the same 401.bzip2 workload, \X{} performs \param{20864} preventive refreshes at 64 banks, leading to a performance overhead of \param{16.21}\%.
Second, \X{} outperforms all four state-of-the-art mechanisms at 16 and 32 banks. \X{} outperforms Hydra, REGA, and PARA at 64 banks and performs very close to (within \param{1.06}\% of the performance of) Graphene.}

\subsection{{Performance Under {Adversarial Workloads}}}
\label{sec:performance-under-attack}
\begin{figure}[h]
    \centering
    \Copy{C3fig}{
    \includegraphics[width=\linewidth]{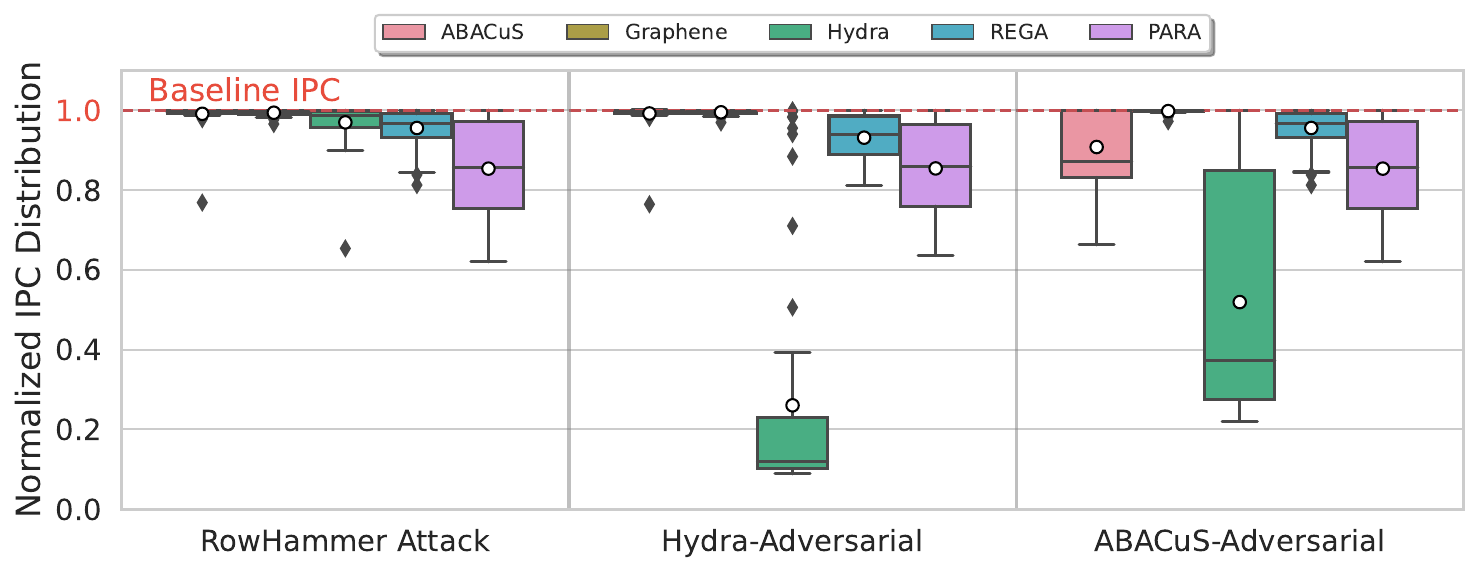}
    \caption{\reviii{Performance comparison for single-core workloads with three different RowHammer access patterns \atbcr{(\gls{nrh} = 500)}}}
    \label{fig:adversarial-patterns}
    }
\end{figure}

\Copy{C3/2}{
\changev{\ref{c:c3}}\X{} securely prevents bitflips under RowHammer attacks~(\secref{sec:security-analysis}). \reviii{We demonstrate that, in a dual-core system, \X{} incurs smaller performance overhead{s} than Hydra, REGA, and PARA for the evaluated single-core workloads on average, while one core in the system executes a traditional RowHammer access pattern (RowHammer Attack) that repeatedly activates 32 rows in each bank in a bank-interleaved manner. We {also} develop two specialized RowHammer access patterns {(which are open source~\cite{self.github})}: Hydra-Adversarial and \X{}-Adversarial. 1)~Hydra-Adversarial exacerbates Hydra's Row Count Cache eviction rate to significantly increase the throughput of Hydra's DRAM read and write requests. 2)~\X{}-Adversarial rapidly increments the spillover counter value to cause frequent refresh cycles (\secref{sec:mechanism-operation}). All three access patterns (RowHammer Attack, Hydra-Adversarial, and \X{}-Adversarial) incur the same, substantially high rate of \gls{act} commands in the memory controller. The memory controller issues an \gls{act} command \emph{every \SI{20}{\nano\second}} while executing each access pattern. \figref{fig:adversarial-patterns} shows the performance impact of \X{} and the state-of-the-art mechanisms on all evaluated single-core workloads in a dual-core system when the second core executes one of the three RowHammer access patterns.}}

\Copy{C3/3}{
\reviii{We make three major observations. First, \X{} induces \emph{only} \param{0.88}\% performance overhead on average across all evaluated workloads when one core executes the RowHammer attack, whereas Graphene, Hydra, REGA, and PARA induce \param{0.61}\%, \param{3.03}\%, \param{4.43}\%, and \param{14.62}\%, respectively. Second, Hydra induces a {large} \param{73.96\%} slowdown, on average across all workloads when one core executes the Hydra-Adversarial access pattern. We attribute this overhead to the high rate of Row Count Cache (RCC) evictions the Hydra-Adversarial access pattern incurs. Hydra {evicts} \param{1.13} RCC entries per last level cache miss on average across all workloads. The memory controller serves an RCC eviction by issuing high-priority \gls{wr} and \gls{rd} DRAM requests (i.e., \gls{wr} and \gls{rd} requests caused by RCC evictions are on the critical path of workload main memory requests). For the same access pattern, \X{} incurs \emph{only} \param{0.79}\% on average across all workloads. Third, \X{} induces \param{9.20}\% performance overhead, on average across all workloads when one core executes the \X{}-Adversarial access pattern. This is because the \X{}-Adversarial access pattern triggers multiple \X{} refresh cycles, during which no memory request can be serviced, while the single-core workload executes. The same access pattern incurs \param{48.08}\% performance overhead on average across all workloads for Hydra.}
}

\begin{figure}[h]
    \centering
    \includegraphics[width=\linewidth]{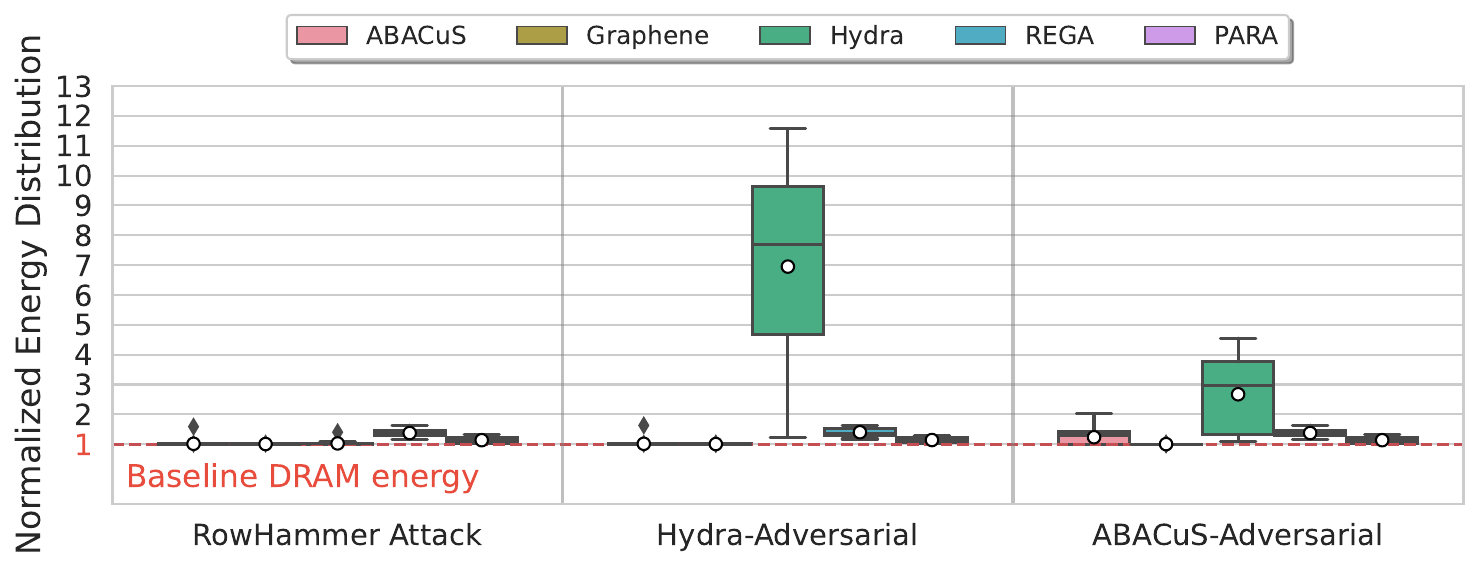}
    \caption{\ext{DRAM energy comparison for single-core workloads with three different RowHammer access patterns \atbcr{(\gls{nrh} = 500)}}}
    \label{fig:adversarial-energy}
\end{figure}

\ext{\figref{fig:adversarial-energy} shows the DRAM energy impact of \X{} and the state-of-the-art mechanisms on all evaluated single-core workloads in a dual-core system when the second core executes one of the three RowHammer access patterns.}

\ext{We make two observations from~\figref{fig:adversarial-energy}. First, \X{} incurs relatively small \param{1.34}\%, \param{1.34}\%, and \param{23.22}\% DRAM energy overheads for the RowHammer Attack, Hydra Adversarial, and \X{}-Adversarial access patterns, respectively. Second, Hydra incurs significant \param{6.95$\times{}\!$} and \param{2.67$\times{}\!$} DRAM energy overheads for the Hydra Adversarial and \X{}-Adversarial access patterns, respectively. We attribute the energy overheads of all mitigation mechanisms to i) preventive refresh operations and ii) increased runtime causing more DRAM background energy consumption.}

{We conclude that ABACuS incurs \omcri{almost-}negligible additional performance overhead on benign workloads when another core executes a traditional RowHammer access pattern in the same system. Specialized adversarial access patterns can exacerbate such overheads by frequently triggering \X{} refresh cycles.}

\subsection{\reviii{Improving \X{}'s Performance Under {Adversarial Workloads}}}
\label{sec:abacus-big}
\begin{figure}[h]
    \centering
    \Copy{C3fig2}{
    \includegraphics[width=\linewidth]{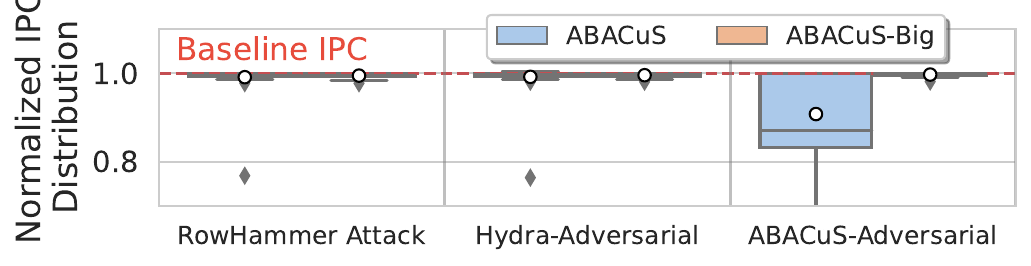}
    \caption{\reviii{Performance comparison \atbcr{of} \X{} and \X{}-Big \atbcr{(\gls{nrh} = 500)}.}}
    \label{fig:adversarial-patterns-big}
    }
\end{figure}

\Copy{C3/4}{
\reviii{A workload may, intentionally (e.g., \X{}-Adversarial) or unintentionally (e.g., gups), rapidly increment the spillover counter's value{, frequently triggering \emph{refresh cycles} where \X{} issues a refresh command to each DRAM row ID in a rank,} and cause substantial performance overheads in an \X{}-based system. To prevent such overheads, a less-area-constrained version of \X{} can remove the spillover counter and implement one shared activation counter (\X{} counter) per DRAM row {ID} {(i.e., \X{}-Big's $N_{entries}$~(Table~\ref{tab:params}) is equal to the number of rows in a DRAM bank)}. We design and evaluate \X{}-Big, which implements one \X{} counter per DRAM row {ID}. The \X{} counter in \X{}-Big is updated in the same way as in \X{}. {\X{}-Big implements as many \X{} counters as there are rows in a bank (i.e., there is a 1-1 mapping between \X{} counters and DRAM row IDs) to keep \emph{precise} track of \emph{every sibling row{'}s} {maximum} activation count and \X{}-Big does \emph{not} need to use a spillover counter.}}  
\figref{fig:adversarial-patterns-big} shows the performance impact of \X{} and \X{}-Big on all evaluated single-core workloads in a dual-core system with the three RowHammer access patterns described earlier in this section.
}

\Copy{C3/5}{
\reviii{We observe that \X{}-Big incurs \emph{only} \param{0.28}\% performance overhead on average across all workloads for the \X{}-Adversarial pattern, whereas \X{} incurs \param{9.20}\% because \X{}-Big does \emph{not} perform any refresh cycles where the memory controller is busy rapidly issuing \gls{ref} commands.} \reviii{We evaluate \X{}-Big's chip area using the methodology described in~\secref{sec:evaluation-area}. \X{}-Big requires 40 bits (8 bits for row activation counter, 32 bits for sibling activation vector) of storage per DRAM row, amounting to 640 KiB on-chip storage at an \gls{nrh} of 500 for 128K DRAM rows. \X{}-Big takes up \param{\SI{0.48}{\milli\metre\squared}} chip area (0.20\% of a high-end Intel Xeon processor's area~\cite{wikichipcascade}).} \reviii{We conclude that \X{}-Big is a high-performance {implementation of the} \X{} design, which system designers that have fewer chip area constraints can choose to implement, that improves system performance under {adversarial workloads and some benign workloads (e.g., gups)} compared to \X{}.}
}

\section{Related Work}
\label{sec:related}
\agy{To our knowledge, \X{} is the first work that mitigates RowHammer efficiently and scalably {at very low \omcri{RowHammer thresholds} (e.g., 125)} without incurring large area, performance, or energy overheads. Sections~\ref{sec:implementation} and \ref{sec:evaluation} already qualitatively and quantitatively {compare} \X{} {to} the most relevant {{state-of-the-art}} mechanisms~\cite{park2020graphene,kim2014flipping,qureshi2022hydra,marazzi2023rega}, {demonstrating \X{}'s benefits}.
This section discusses other RowHammer mitigation mechanisms.}

\noindent
\textbf{Hardware-based Mitigation Mechanisms.} Many prior work\agy{s}~\cite{kim2014flipping, son2017making, you2019mrloc, lee2019twice, park2020graphene,yaglikci2021blockhammer , greenfield2012throttling, qureshi2022hydra, saileshwar2022randomized,kim2022mithril, yaglikci2022hira, saxena2022aqua, marazzi2022protrr,hassan2022case, devaux2021method, woo2022scalable, kim2015architectural, wang2021discreet, bennett2021panopticon, joardar2022learning, joardar2022machine, zhou2022ltpim, lee2021cryoguard, saroiu2022configure, kang2020cattwo, seyedzadeh2017cbt, seyedzadeh2018cbt, hassan2019crow} propose hardware-based mitigation mechanisms to prevent RowHammer bitflips. {We classify these into three main categories. 1) Probabilistic preventive refresh (PPR) mechanisms~\cite{kim2014flipping, you2019mrloc, son2017making, wang2021discreet, yaglikci2022hira, joardar2022learning, joardar2022machine, zhou2022ltpim} preventively refresh victim rows based on a probability. PPR mechanisms incur impractical performance overheads at very low RowHammer thresholds as they perform many unnecessary preventive refresh operations. A recent work~\cite{saroiu2022configure} proposes a new methodology for configuring PPRs. 2) Deterministic preventive refresh (DPR) mechanisms~\cite{kim2015architectural, seyedzadeh2017cbt, seyedzadeh2018cbt, kang2020cattwo, lee2019twice, saileshwar2022randomized, saxena2022aqua, kim2022mithril, marazzi2022protrr, park2020graphene, woo2022scalable, bennett2021panopticon, hassan2019crow, lee2021cryoguard, hassan2022case, devaux2021method} track activation counts of aggressor rows and preventively refresh victim rows. DPR mechanisms incur less performance overhead than PPR mechanisms (from fewer unnecessary preventive refresh operations) at the cost of larger chip area overhead to store aggressor row activation counters. and 3) Deterministic aggressor row access throttling (DAT) mechanisms~\cite{yaglikci2021blockhammer,kim2014flipping,greenfield2012throttling} track activation counts of aggressor rows and preventively block memory accesses to aggressor rows. \emph{DAT mechanisms} incur average system performance and total chip area overheads similar to DPR mechanisms~\cite{yaglikci2021blockhammer}. However, existing \emph{DAT mechanisms} can incur delays in the order of microseconds on memory demand requests (e.g., load instructions)~\cite{saxena2022aqua,qureshi2022hydra,yaglikci2021blockhammer}.}

\noindent
\yct{\textbf{Software-based Mitigation Mechanisms.} \agy{Many works~\cite{konoth2018zebram, van2018guardion, brasser2017can, bock2019riprh, aweke2016anvil, zhang2022softtrr, enomoto2022efficient}
propose software-based mitigation mechanisms to avoid hardware modifications. Unfortunately, it is \emph{not} possible for these mechanisms to monitor \omcri{\emph{all}} memory requests, and thus most of these mechanisms have already been defeated by recent attacks~\cite{qiao2016new, gruss2016rowhammer, gruss2018another, cojocar2019eccploit, zhang2019telehammer, kwong2020rambleed, zhang2020pthammer}.}
}

\noindent
\ste{\textbf{Integrity-based Mitigation Mechanisms.} 
\agy{Several works~\cite{juffinger2023csi, fakhrzadehgan2022safeguard, qureshi2021rethinking, manzhosov2022revisiting} propose integrity check mechanisms to detect and correct bitflips that may have been induced by RowHammer. Unfortunately, it is either not possible\omcri{, very} difficult, or prohibitively expensive to correct all possible RowHammer bitflips using these mechanisms. However, these mechanisms can be combined with \X{} to improve overall system reliability \revcommon{and future work could demonstrate the benefits of combining them with \X{}}.}
}

\glsresetall

\noindent
\yct{\textbf{RowHammer Mitigation in Commodity Chips.}}
\Copy{C5/1}{
\changev{\ref{c:c5}}\agy{DRAM manufacturers employ RowHammer mitigation mechanisms, \revcommon{commonly referred to as target row refresh (TRR)}, in commodity DRAM chips~\cite{jedec2017ddr4,jedec2020ddr5} without publicly documenting their detailed designs.} \revv{These mechanisms typically do \emph{not} induce any performance overhead because they take action (e.g., refresh a victim row) when the DRAM chip is busy performing a periodic refresh operation (i.e., their victim row refresh latency is hidden by the latency of performing a periodic refresh operation). However,} recent studies experimentally demonstrate that specialized adversarial access patterns can defeat {some of} these mechanisms~\cite{hassan2021utrr,frigo2020trrespass,van2016drammer,jattke2022blacksmith,deridder2021smash,saroiu2022price}. {A recent work~\cite{naseredini2022alarm} develops a tool that can automatically infer parameters of TRR mechanisms.} {Appendix~\ref{appendix:trr-security-analysis} analyzes the security guarantees of a widely-used TRR mechanism, whose inner workings were uncovered by~\cite{hassan2021utrr}, and shows that this TRR mechanism \emph{cannot} securely prevent RowHammer bitflips at experimentally demonstrated (e.g., 4.8K~\cite{kim2020revisiting}) RowHammer thresholds.} {Recent works from industry design new probabilistic in-DRAM RowHammer mitigation mechanisms~\cite{kim2023ddr5,hong2023dsac}. Unfortunately, these mechanisms cannot \omcri{or are not proven to} deterministically prevent all RowHammer bitflips.}
}

\noindent
{\textbf{Device-level Mechanisms for Mitigating RowHammer.} Several prior works~\cite{yang2016suppression,gomez2016dummy,han2021surround,ryu2017overcoming} design new DRAM cells or new DRAM arrays with improved RowHammer resilience. Unfortunately, these works alone \emph{cannot} completely prevent RowHammer bitflips \atbcr{(but could effectively increase the RowHammer thresholds)}. However, they can be used together with other hardware- or software-based mitigation techniques to mitigate RowHammer.}

\section{Conclusion}
\label{sec:conclusion}

We introduced a new RowHammer mitigation mechanism that prevents RowHammer bitflips at low area, performance, and energy overheads for modern and future DRAM chips that are {very} vulnerable to RowHammer {(e.g., with hammer counts as low as 125 inducing bitflips)}. Compared
to existing RowHammer mitigation mechanisms, our {technique,} all-bank activation counters for scalable RowHammer mitigation (\X{}) {technique} incurs significantly smaller area, performance, and DRAM energy overheads for modern and future DRAM chips. Our technique achieves this by sharing activation counters of rows that has the same row \gf{ID} in different banks. While \X{} efficiently and securely prevents RowHammer bitflips, it also scales well with worsening RowHammer vulnerability {down to \gls{nrh} = 125}.

\section*{Acknowledgments}
{We thank our shepherd and the anonymous reviewers of USENIX Security 2023. \ext{We thank Rakesh Kumar for their valuable feedback.} We thank the SAFARI Research Group members for feedback and the stimulating intellectual environment. We acknowledge the generous gift funding provided by our industrial partners (especially Google, Huawei, Intel, Microsoft, VMware), which has been instrumental in enabling our decade-long research on read disturbance in DRAM and memory systems. This work was in part supported by a Google Security and Privacy Research Award and the Microsoft Swiss Joint Research Center.}


\bibliographystyle{unsrt}

\bibliography{main}
\normalsize
\appendix
\section*{\LARGE{Appendix}}

\section{\X{} Security Analysis}
\label{appendix:security-proof}

We explain how \X{} maintains the maximum activation count \atbcr{among all sibling rows (described in~\secref{sec:mechanism})} in the row activation counters by showing that Invariant~\ref{th:max-count} holds after the row activation counter is updated by \X{} upon activation of a DRAM row.

\textbf{Analysis Overview.} Invariant~\ref{th:max-count} formally defines the property of the value stored in a row activation counter in terms of the \emph{actual} activation count of sibling DRAM rows (i.e., the absolute number of activations each sibling DRAM row receives, regardless of the row activation counter's value) in a \gls{trefw}. The spillover counter already stores a value greater than or equal to a row's activation count for DRAM rows that are \emph{not} tracked by any \X{} counter. \atbcr{We provide a simple explanation and refer the reader to~\cite{park2020graphene} for comprehensive proof of why the spillover counter's value is greater than or equal to a non-tracked row's activation count. Initially and after periodic reset, all \X{} counter values and the spillover counter value are all zero. Any activated row ID is tracked in one of the \X{} counters until no unassigned (to a row ID) \X{} counters are left. When there are no unassigned \X{} counters left, the spillover counter is still 0, and the spillover counter is incremented with each activation to an unmapped (i.e., non-tracked) row ID until the spillover counter value equals the minimum (there may be multiple of such counters) \X{} counter value. If only one unmapped row ID receives all these activations, the spillover counter value equals the activation count of this unmapped row. Otherwise, the spillover counter value exceeds the activation counts of multiple unmapped rows.} 

\begin{mytheo}{}{max-count}
  \footnotesize 
  Let $ACT\_COUNT(bank)(row ID)$ denote the \emph{actual} activation count of a $row$ \omcri{with row ID} in a $bank$. If a $row$ is tracked by an \X{} counter, the row activation counter (RAC) corresponding to this row is always greater than or equal to the actual activation count of the $row$ \omcri{with the same row ID} in \emph{any} of the banks. That is, \\
  $\forall{}b'\in{}Banks$ with sibling row $r'$ of $r$. $RAC(r)\geq{} ACT\_COUNT(b')(r')$

\end{mytheo}

\noindent\textbf{Proof:} By induction on the actual activation count of row $r$ in bank $b$, tracked by $RAC(r)$.

\noindent\textbf{Base Case:} When an $RAC$ counter starts tracking a row $r$ in bank $b$ for the first time, the following holds:

\begin{itemize}
    \item $RAC(r)\geq ACT\_COUNT(b)(r)$
    \item $SAV(b)(r)$ is set. Other $SAV$ bits are zero.
\end{itemize}

\noindent\textbf{Induction Hypothesis:} Assume that invariant holds for any row $r'$ in bank $b'$ which are tracked by an \X{} counter.

\noindent\textbf{Step Case:} Let $r'$ be an arbitrary sibling row of $r$ in bank $b'$. Note that $RAC(r') = RAC(r)$ by definition of $RAC$. Assume that such an $r'$ is activated. We distinguish between two cases.

\noindent\textbf{Case 1:} $\bm{SAV(b')(r')}$ \textbf{\textit{is not set.}} In this subcase, before activating $r'$, we have $RAC(r')>ACT\_COUNT(b')(r')$, since otherwise $SAV(b')(r')$ would have been set. Therefore, after activating $r'$ we have $RAC(r')\geq{} ACT\_COUNT(b')(r')$ and $SAV(b')(r')$ is set.

\noindent\textbf{Case 2:} $\bm{SAV(b')(r')}$ \textbf{\textit{is set}}. In this subcase, before activating $r'$, we have $RAC(r')\geq{}ACT\_COUNT(b')(r')$. Hence, after activating $r'$, the actual activation count of $r'$ increases by one, and $RAC(r')$ is incremented. The $SAV(b')(r')$ remains set, while other $SAV$ bits are reset. Thus, $RAC(r')\geq{}ACT\_COUNT(b')(r')$ still holds, satisfying the invariant.

\section{Single Core Performance and Energy Results for Each Workload}
\label{appendix:single-core-extended}

Table~\ref{table:rbmpki} shows the row buffer misses per kilo instruction (RBMPKI) of each tested workload.

\begin{table}[!ht]
\centering
\caption{Row buffer misses per kilo instruction (RBMPKI) of evaluated single-core workloads}
\resizebox{\linewidth}{!}{
\begin{tabular}{|c|l|l||c|l|l|}
\midrule
\textbf{RBMPKI Class} & \textbf{Workload}      & \textbf{RBMPKI}  & \textbf{RBMPKI Class} & \textbf{Workload}         & \textbf{RBMPKI}    \\
\midrule
\multirow{27}{*}{LOW\_RBMPKI}    & h264\_encode  & 0.00019 & \multirow{17}{*}{MED\_RBMPKI}    & stream\_10.trace & 2.87468   \\
                                 & 511.povray    & 0.00187 &                                  & tpcc64           & 3.04086   \\
                                 & 481.wrf       & 0.00392 &                                  & ycsb\_aserver    & 3.12505   \\
                                 & 541.leela     & 0.00444 &                                  & 557.xz           & 3.56674   \\
                                 & 538.imagick   & 0.01055 &                                  & 482.sphinx3      & 3.86501   \\
                                 & 444.namd      & 0.01744 &                                  & jp2\_decode      & 3.92657   \\
                                 & 447.dealII    & 0.01939 &                                  & 505.mcf          & 4.05419   \\
                                 & 464.h264ref   & 0.02927 &                                  & wc\_8443         & 4.42091   \\
                                 & 456.hmmer     & 0.08931 &                                  & wc\_map0         & 4.42437   \\
                                 & 403.gcc       & 0.17105 &                                  & 436.cactusADM    & 5.10635   \\
                                 & 526.blender   & 0.18119 &                                  & 471.omnetpp      & 6.56474   \\
                                 & 544.nab       & 0.25256 &                                  & 473.astar        & 6.74379   \\
                                 & 525.x264      & 0.35040 &                                  & jp2\_encode      & 6.82960   \\
                                 & 508.namd      & 0.37629 &                                  & tpch17           & 7.26692   \\
                                 & grep\_map0    & 0.40331 &                                  & 483.xalancbmk    & 7.58744   \\
                                 & 531.deepsjeng & 0.40342 &                                  & 462.libquantum   & 8.86578   \\
                                 & 458.sjeng     & 0.51727 &                                  & tpch2            & 9.81832   \\   
\cline{4-6}
\cline{4-6}
                                 & 435.gromacs   & 0.54579 & \multirow{15}{*}{HIGH\_RBMPKI}   & 433.milc         & 11.43486  \\
                                 & 445.gobmk     & 0.55018 &                                  & 520.omnetpp      & 12.40926  \\
                                 & 401.bzip2     & 0.58087 &                                  & 437.leslie3d     & 16.64157  \\
                                 & 507.cactuBSSN & 0.77452 &                                  & 450.soplex       & 17.09248  \\
                                 & 502.gcc       & 1.03991 &                                  & 459.GemsFDTD     & 19.25577  \\
                                 & ycsb\_abgsave & 1.19071 &                                  & 549.fotonik3d    & 21.44822  \\
                                 & tpch6         & 1.21633 &                                  & 434.zeusmp       & 22.23071  \\
                                 & 500.perlbench & 1.54678 &                                  & 519.lbm          & 36.63709  \\
                                 & 523.xalancbmk & 1.63728 &                                  & 470.lbm          & 41.76373  \\
                                 & ycsb\_dserver & 1.79053 &                                  & 429.mcf          & 71.59797  \\
\cline{1-3}
\cline{1-3}
\multirow{5}{*}{MED\_RBMPKI}     & ycsb\_cserver & 2.12396 &                                  & gups & 87.37615  \\
                                 & 510.parest    & 2.17283 &                                  & h264\_decode     & 204.46196 \\
                                 & ycsb\_bserver & 2.21749 &                                  & bfs\_ny          & 219.56454 \\
                                 & ycsb\_eserver & 2.63139 &                                  & bfs\_cm2003      & 219.78414 \\
                                 &               &         &                                  & bfs\_dblp        & 219.79677 \\
\toprule
\end{tabular}
}
\label{table:rbmpki}
\end{table}

\begin{figure*}[!th]
    \centering
    \includegraphics[width=0.93\textwidth]{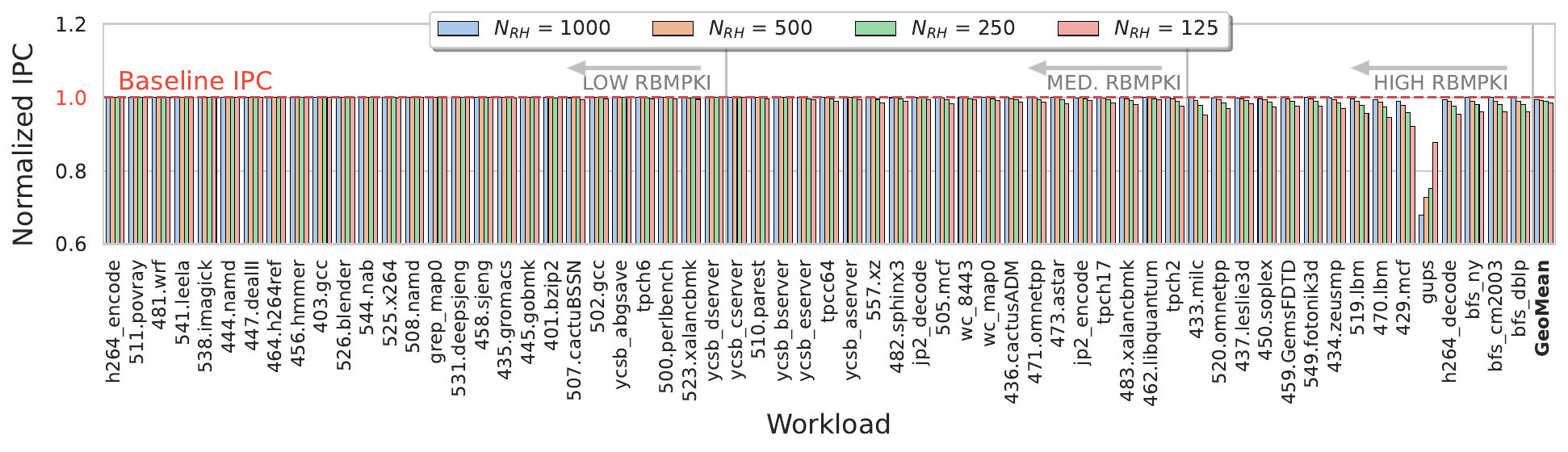}
    \caption{\omcri{Normalized} performance of single-core applications for four different RowHammer thresholds (higher is better).}
    \label{fig:perf_single_core}
\end{figure*}

\begin{figure*}[!th]
    \centering
    \includegraphics[width=0.93\textwidth]{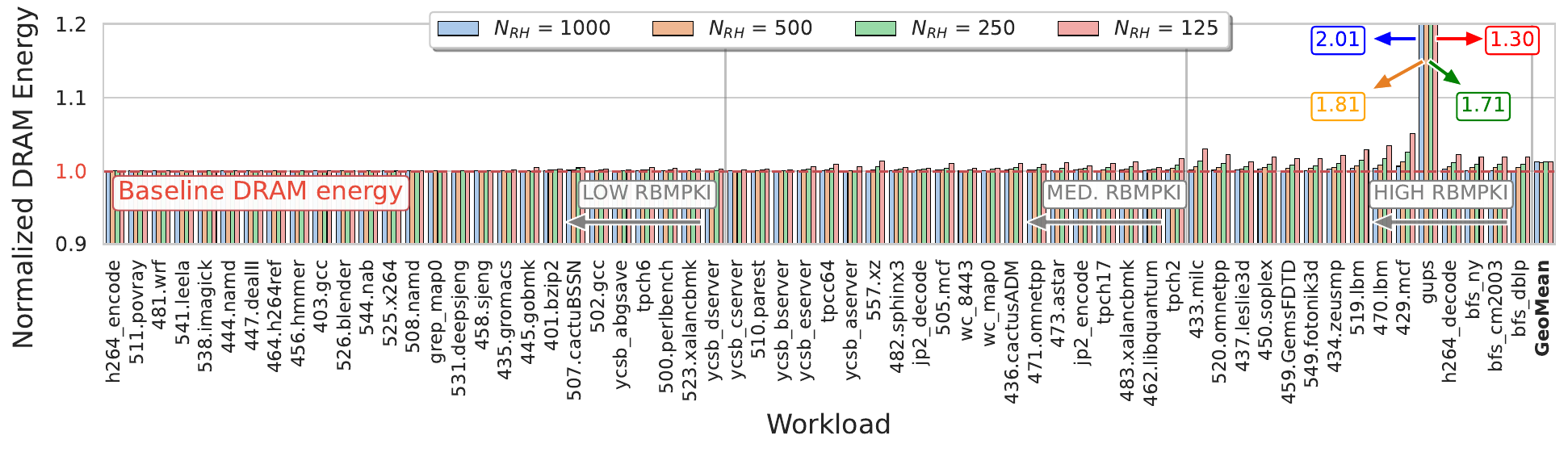}
    \caption{\omcri{Normalized} DRAM energy of single-core applications for four different RowHammer thresholds (lower is better).}
    \label{fig:energy_single_core}
\end{figure*}

\figref{fig:perf_single_core} and~\figref{fig:energy_single_core} plot performance and DRAM energy normalized to baseline for each workload, respectively. \omcri{We sort the workloads on the x-axis in increasing row buffer misses per kilo instruction (RBMPKI) from left to right.} \atbcr{We label the normalized DRAM energy consumption of the gups workload at \gls{nrh} = 1000, 500, 250, and 100 using blue, orange, green, and red colors in~\figref{fig:energy_single_core}, respectively.}

\section{Sensitivity Analyses}

\subsection{Effect of Blast Radius}
\label{sec:eval-blast-radius}

\begin{figure}[!h]
    \centering
    \includegraphics[width=0.95\linewidth]{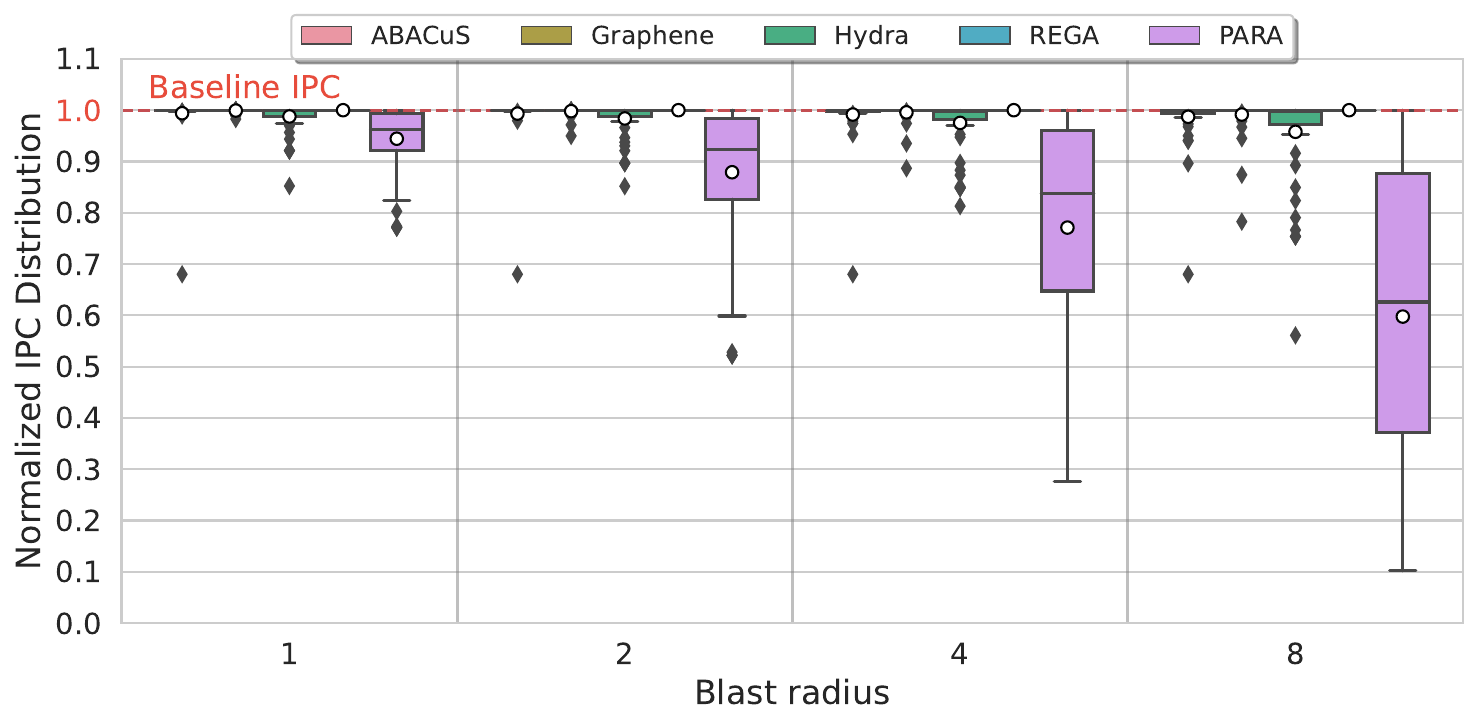}
    \caption{\ext{Normalized performance (distribution) of single-core applications with varying blast radius for ABACuS and four state-of-the-art mechanisms. We perform this analysis for a RowHammer threshold of 1000.}}
    \label{fig:sensitivity-blast}
\end{figure}

\ext{\figref{fig:sensitivity-blast} shows the performance of \X{} and four state-of-the-art mechanisms' performance overheads over four blast radius (see~\secref{sec:blast}) values normalized to baseline (for \gls{nrh} = 1000).}

\ext{From~\figref{fig:sensitivity-blast}, we make two key observations. First, \X{} has relatively small \param{0.58}\%, \param{0.66}\%, \param{0.85}\%, and \param{1.28}\% average performance overheads for blast radius = 1, 2, 4, and 8, respectively. Second, as blast radius increases, all mitigation mechanisms except REGA induce increasing performance overheads (at \gls{nrh} = 1000). This is because higher blast radii necessitate more victim row refreshes, i.e., more activate and precharge commands on the critical path of main memory demand requests. REGA's performance overhead does \emph{not} increase with increasing blast radius because the increase in blast radius (from 1 to 8) does \emph{not} require an increase in \gls{trc}. However, REGA has prohibitive performance overheads for smaller \gls{nrh} values, as shown in~\figref{fig:perf_comparison_single_core}, even at a blast radius of 1.}

\subsection{\ext{Effect of Address Mapping}}
\label{appendix:mapping}

\begin{figure*}[!th]
    \centering
    \includegraphics[width=0.93\textwidth]{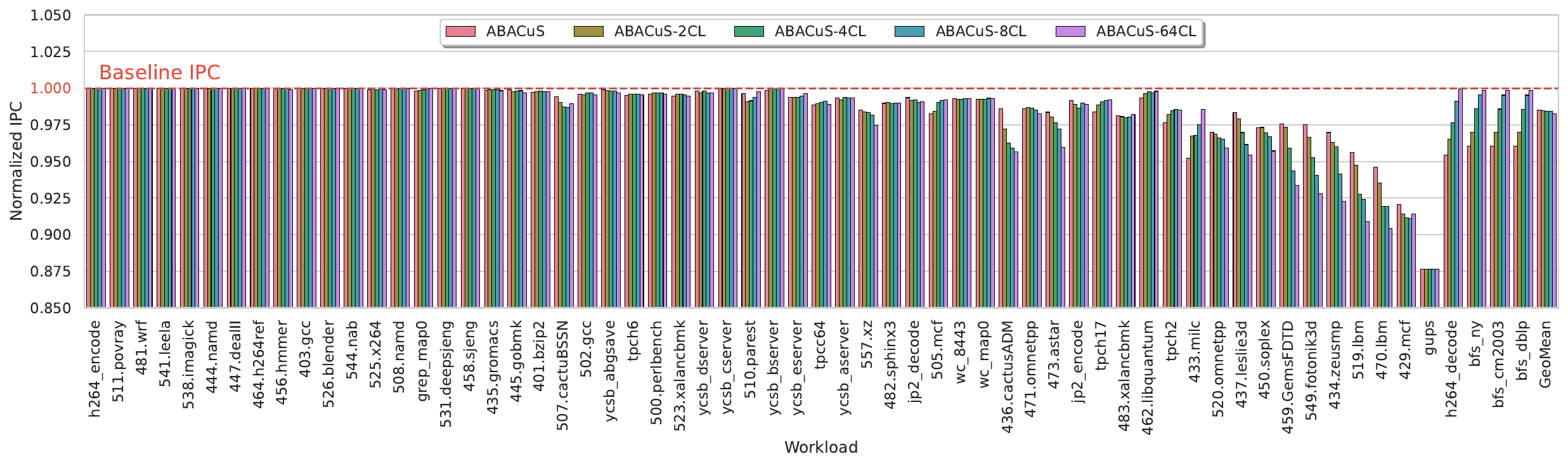}
    \caption{\ext{\omcri{Normalized} performance of single-core applications for four different physical address to DRAM address mappings (\gls{nrh} = 125).}}
    \label{fig:sensitivity-mapping}
\end{figure*}

\ext{\figref{fig:sensitivity-mapping} shows performance of all 62 single-core workloads for four different physical address to DRAM address mappings at \gls{nrh} = 125 for \X{}, normalized to the baseline system that implements the same address mapping as \X{}. \X{} with the default address mapping (depicted as \X{}) interleaves consecutive cache blocks in the physical address space between different DRAM banks. \X{}-2CL interleaves two adjacent cache blocks between different DRAM banks and puts the two cache blocks next to each other in the same row. \X{}-4CL, \X{}-8CL, and \X{}-64CL interleave four, eight, and 64 (one memory page worth of) adjacent cache blocks between different DRAM banks and put the four, eight, and 64 adjacent cache blocks next to each other in the same row, respectively.}

\ext{We make two major observations from~\figref{fig:sensitivity-mapping}. First, \X{}'s average performance overhead marginally changes with the address mapping. \X{}, \X{}-2CL, \X{}-4CL, and \X{}-8CL incur 1.45\%, 1.48\%, 1.53\%, and 1.55\% performance overheads on average across all 62 workloads, respectively. Second, address mappings that put more adjacent cache blocks into the same DRAM row can increase or reduce \X{}'s performance overheads. For example, \X{} incurs 7.95\% and \X{}-8CL incurs 8.88\% overhead on 429.mcf, whereas \X{} incurs 4.57\% and \X{}-8CL incurs 0.90\% overhead on h264\_decode. 
Some workloads (e.g., 429.mcf) increase \X{} counters more rapidly (and perform more preventive refresh operations) whereas others (e.g., h264\_decode) increase \X{} counters more slowly (and perform fewer preventive refresh operations) when the system employs an address mapping that puts more adjacent cache lines into the same row. We investigate h264\_decode's performance in more detail. This workload's row buffer hit rate significantly increases as we interleave more adjacent cache blocks between DRAM banks. Therefore, with increasing number of interleaved cache blocks, the memory controller issues fewer activate commands and thus \X{} counters reach the preventive refresh threshold more infrequently for this workload.}

\ext{We conclude that address mapping has a marginal affect on \X{}'s mean performance across all evaluated single-core workloads. No single address mapping minimizes \X{}'s performance overheads for all tested workloads.}

\subsection{\ext{Heterogeneous 8-Core Workload Performance}}
\label{appendix:heterogeneous-workloads}

\begin{figure}[!h]
    \centering
    \includegraphics[width=0.95\linewidth]{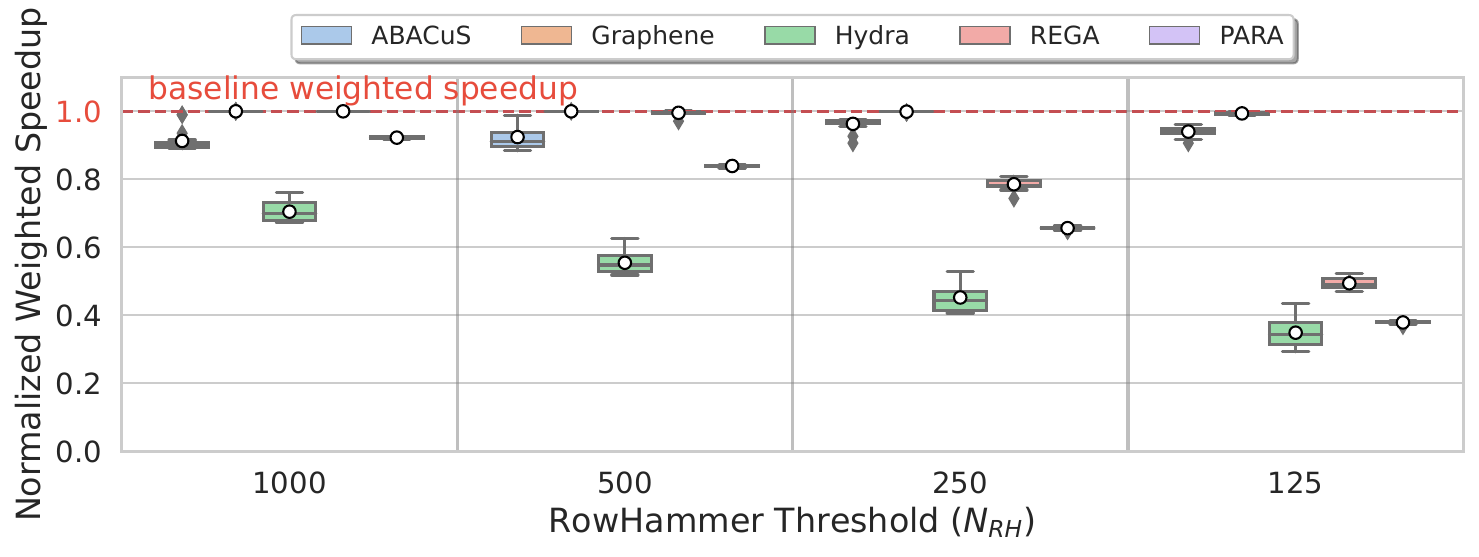}
    \caption{\ext{Normalized performance (distribution) of heterogeneous 8-core applications with varying RowHammer Threshold for ABACuS and four state-of-the-art mechanisms.}}
    \label{fig:heterogeneous-workloads}
\end{figure}

\ext{We create 16 heterogeneous 8-core workloads by randomly sampling the HIGH RBMPKI workloads listed in Table~\ref{table:rbmpki}. Figure~\ref{fig:heterogeneous-workloads} shows the performance (in terms of weighted speedup) of \X{} and four other state-of-the-art mechanisms normalized to the baseline. Our observations for the heterogeneous 8-core workloads are in line with those for the homogeneous 8-core workloads presented in~\secref{sec:evaluation-performance-energy}.}

\section{TRR Security Analysis}
\label{appendix:trr-security-analysis}

\revv{To demonstrate the minimum \gls{nrh} that known TRR mechanisms can securely prevent RowHammer bitflips at, we evaluate a widely-adopted TRR mechanism (whose inner workings are uncovered by~\cite{hassan2021utrr}). This mechanism corresponds to the ones used by Vendor \param{\ste{A}} in Table \param{\ste{1}} in~\cite{hassan2021utrr}.}

\begin{figure}[!h]
    \centering
    \Copy{C5fig}{
    \includegraphics[width=\linewidth]{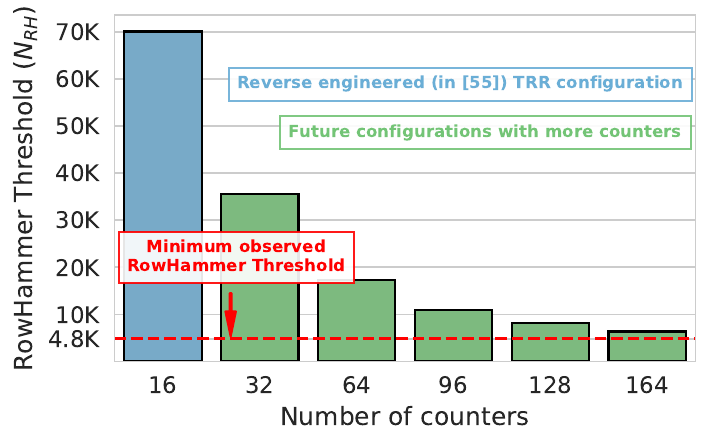}
    \caption{\revv{Minimum RowHammer threshold (\gls{nrh}) values that Vendor A's TRR in~\cite{hassan2021utrr} can prevent RowHammer bitflips \atbcr{for} different number of counters.}}
    \label{fig:trr-thresholds}
    }
\end{figure}

\Copy{C5/2}{
\revv{The TRR mechanism likely adopts the Misra-Gries algorithm and implements 16 counters to track aggressor rows. We assume that the DRAM chip can refresh all 16 tracked aggressor rows with every periodic refresh command. We find the maximum number of aggressor row activations that a carefully-engineered \omcri{many-sided} RowHammer access pattern (based on the access patterns described in~\cite{hassan2021utrr}) perform \emph{before} TRR refreshes victim rows. This access pattern tricks TRR into \emph{not} tracking a \emph{real} aggressor row, by repeatedly accessing multiple \emph{dummy} rows (e.g., 16 such rows if there are 16 aggressor row activation counters) more times than the real aggressor rows. \figref{fig:trr-thresholds} shows the maximum aggressor row activation counts (y-axis) that the access pattern achieves when the DRAM chip employs the reverse engineered TRR mechanism as is (\atbcr{the blue} leftmost bar) and the chip employs future versions of the TRR mechanism that implement more aggressor row activation counters (\atbcr{green} bars toward right), where we depict the number of activation counters on the x-axis.}
}

\Copy{C5/3}{
\revv{We observe that even a substantially-scaled version of the TRR mechanism, which implements more than 10$\times{}$ the counters \omcri{(rightmost bar in the figure)} as the reverse engineered one, \emph{cannot} prevent RowHammer bitflips at an \gls{nrh} value \ste{of $4.8$K} that prior work experimentally demonstrated real chips to exhibit~\cite{kim2020revisiting}. We conclude that \omcri{existing} TRR \omcri{mechanisms implemented in DRAM chips~\cite{hassan2021utrr,frigo2020trrespass}} \emph{cannot} prevent RowHammer bitflips at today's (e.g., $4.8$K) or futuristic \gls{nrh} values \ste{(e.g., $125$)}.}}


\section{Artifact Appendix}


\subsection{Abstract}

Our artifact contains the data, source code, and scripts needed to reproduce our results. We provide: 1)~the source code of our simulation infrastructure based on Ramulator and CACTI, and 2)~all evaluated workloads' memory access traces and all major evaluation results. We provide Python scripts and Jupyter Notebooks to analyze and plot the results.

\subsection{Description \& Requirements}

\emph{We strongly recommend using Slurm for running experiments in bulk.}

\subsubsection{Security, privacy, and ethical concerns}

None.

\subsubsection{How to access}

The source code and analysis scripts are provided in our open source repository at the following stable reference \url{https://github.com/CMU-SAFARI/ABACuS/tree/7491a667fd1a667b556ef81a8eaa035f69461644}.
We provide all evaluated workloads' memory access traces and all evaluation results at \url{https://zenodo.org/doi/10.5281/zenodo.10575682}.

\subsubsection{Hardware dependencies}

We recommend using a PC with 32 GiB of main memory. Approximately 50 GiB of disk space is needed to store intermediate and final experimental results.


\subsubsection{Software dependencies}

\texttt{Podman} (we have tested Podman version 3.4.4 on Ubuntu 22.04.1) and \texttt{git}. 






\subsubsection{Benchmarks}

We use workload memory traces collected from SPEC2006, SPEC2017, TPC, MediaBench, and YCSB benchmark suites. 
The used memory traces are available at \url{https://zenodo.org/doi/10.5281/zenodo.10575682}.
A provided script will download and extract the traces.

\subsection{Set-up}

\subsubsection{Installation}

Clone the git repository using \texttt{\$~git clone -b usenix24-ae git@github.com:CMU-SAFARI/ABACuS.git}.

\subsubsection{Basic Test}

Run \texttt{./simple\_test\_podman.sh}. This will download and extract all workload execution traces as an intermediate step, which might take around 20 minutes. Expected results are the same as above.


\subsection{Evaluation workflow}

\subsubsection{Major Claims}


\begin{itemize}
\itemindent=15pt

    \item[(C1):] Modern memory-intensive workloads and existing RowHammer attacks activate DRAM rows with the same row address in multiple DRAM banks (i.e., sibling rows) at around the same time. A single shared activation counter, which stores the highest activation count among the activation counts of all sibling rows, can {reasonably} accurately represent the activation count of all sibling rows. This property of the shared activation counter becomes stronger as the RowHammer threshold reduces. This is proven by the experiment (E1) described in Section 3.1 whose results are illustrated in Figures 2 and 3. 

    \item[(C2):] \X{} induces small system performance and DRAM energy overheads on average across all tested single-core and multi-core workloads for RowHammer threshold values of 1000, 500, 250, and 125. \X{}'s performance and DRAM energy overheads are closer to the most-performance-efficient state-of-the-art mechanism. \X{} outperforms and consumes less DRAM energy than the most-area-efficient state-of-the-art (counter-based) mechanism. This is proven by the experiment (E2) described in Sections 8 and 9 whose results are illustrated in Figures 7, 8, 9, 10, 11, and 12.

    \item[(C3):] At the RowHammer threshold (nRH) {of 125}, \X{} {performs very similarly to the best prior performance- and energy-efficient RowHammer mitigation mechanism while requiring 22.72$\times{}$ smaller chip area. This is proven by the experiment (E3) described in Section 7.1 whose results are illustrated in Table 1.}

\end{itemize}
\setstretch{0.98}
\subsubsection{Experiments}

    \textbf{(E1 and E2):} \textit{[Ramulator simulations] [15 human-minutes + 77 compute-hours (assuming approx. 1280 cores available for running tasks in parallel) + 50GB disk]:
    Execute Ramulator simulations to generate data supporting C1 and C2. Plot all figures that prove C1 and C2.}

    \begin{enumerate}
        \item Execute \texttt{\$~./run\_artifact\_with\_podman.sh --personalcomputer}. This will launch all Ramulator simulation jobs. Wait for simulations to end.
        \item Run \texttt{\$~./create\_figures\_with\_podman.sh}. The script takes approximately 5 hours to execute (a large fraction of the execution time is spent on one-time preprocessing of the data used for Figures 2 and 3).
    \end{enumerate}
    A PDF of every figure proving C1 and C2 is created in \texttt{scripts/ae\_scripts/}.

    \noindent
    \textbf{(E3):} \textit{[CACTI simulations] [5 human-minutes + 1 compute-minute + 10MB disk]:
    Run CACTI simulations to generate chip area estimation results.}

    \begin{enumerate}
        \item Run \texttt{\$~./area\_results\_with\_podman.sh}. 

    \end{enumerate}
    The area cost of ABACuS, Graphene, and Hydra is printed (along with all data used to fill the cells of Table 1). The ratio of Graphene's ``Total Area'' at ``nRH:125'' (5.68$mm^2$) to ABACuS's ``Total Area'' at ``nRH:125'' (0.25$mm^2$) is 22.72$\times{}$.

\subsection{Version}
Based on the LaTeX template for Artifact Evaluation V20231005. Submission,
reviewing and badging methodology followed for the evaluation of this artifact
can be found at \url{https://secartifacts.github.io/usenixsec2024/}.


\end{document}